\documentclass[useAMS,usenatbib]{mnras}
\usepackage{graphicx}
\usepackage{txfonts}
\usepackage{color} 

\title[Spiral arms in CALIFA galaxies]
      {Spiral arms in CALIFA galaxies traced by non-circular velocities, abundances, and extinctions}

\author[F.~Sakhibov et al.]
       {F.~Sakhibov$^{1}$,
        I.~A.~Zinchenko$^{2,3}$,
        L.~S.~Pilyugin$^{2,3}$,
        E.~K.~Grebel$^{3}$,
        A.~Just$^{3}$,
        J.~M.~V\'{i}lchez$^{4}$
\\
$^{1}$ University of Applied Sciences of Mittelhessen, Campus Friedberg,
        Department of Mathematics, Natural Sciences and Data Processing, \\
        Wilhelm-Leuschner-Stra\ss e 13, 61169 Friedberg, Germany \\
 $^{2}$ Main Astronomical Observatory of National Academy of Sciences of Ukraine,
           27 Zabolotnogo str., 03680 Kiev, Ukraine \\
 $^{3}$ Astronomisches Rechen-Institut, Zentrum f\"ur Astronomie der Universit\"at 
       Heidelberg, M\"onchhofstr.\ 12--14, 69120 Heidelberg, Germany \\   
 $^{4}$ Instituto de Astrof\'{i}sica de Andaluc\'{i}a, CSIC, Apdo, 3004, 18080 Granada, Spain \\    
             }

\date{Accepted 2017 Month 00. Received 2017 Month 00; in original form 2017 March 09}

\begin{document}

\maketitle

\begin{abstract}
We derive maps of the observed velocity of ionized gas, the oxygen
abundance, and the extinction (Balmer decrement) across the area of
the four spiral galaxies NGC~36, NGC~180, NGC~6063, and NGC~7653 from
integral field spectroscopy obtained by the Calar Alto Legacy Integral
Field Area (CALIFA) survey.  We searched for spiral arms through
Fourier analysis of the spatial distribution of three tracers
(non-circular motion, enhancement of the oxygen abundance, and of the
extinction) in the discs of our target galaxies. The spiral arms
(two-armed logarithmic spirals in the deprojected map) are shown in
each target galaxy for each tracer considered.  The pitch angles of
the spiral arms in a given galaxy obtained with the three different
tracers are close to each other.  The enhancement of the oxygen
abundance in the spiral arms as compared to the abundance in the
interarm regions at a given galactocentric distance is small; within a
few per cent. We identified a metallicity gradient in our target galaxies.
Both barred galaxies in our sample show flatter gradients than the two
galaxies without bars.  Galactic inclination, position angle of the major
axis, and the rotation curve were also obtained for each target galaxy
using the Fourier analysis of the two-dimensional velocity map.
\end{abstract}


\section{Introduction}
\label{sect:intro}

Spiral arms are a prominent feature of late-type regular disc
galaxies.  The spirals arms in galaxies have been investigated via the
analysis of the spatial distribution of different tracers such as the
positions of H\,{\sc ii} regions \citep[e.g.,][]{kalnajs1975,
considere1982}, the light distribution at a given wavelength
\citep[e.g.,][]{Iye1982, Elmegreen1984, considere1988, Eskridge2002},
the velocity of gas clouds \citep[e.g.,][]{rots1975, sakhibov1987,
sakhibov1989, sakhibov1990, sakhibov2004,canzian1993, fridman2001a,
fridman2001b}, the dust distribution \citep[e.g.,][]{grosbol1999},
magnetic fields \citep[e.g.,][]{Vallee1995, frick2016}, etc.
 
The investigation of spiral arms requires a detailed map of the chosen
tracers across the galactic disc.  Spiral arms appear as a
deviation from the axisymmetric distribution of the tracer (e.g.,
velocity, abundance, etc.).  Since the amplitude of the deviation is
small, high-quality measurements of the tracer are required in order
to reveal the spiral arm.  Integral field spectroscopy of galaxies
provides information about the spatial distribution of different
tracers simultaneously and allows us to compare between parameters of
the spiral arms determined through different tracers. This is
particularly useful when examining the link between the kinematical
and other properties of spiral arms.

Observations with integral field spectroscopy were carried out for a
large sample of galaxies in the framework of the ``Calar Alto Legacy
Integral Field Area Survey'' \citep[CALIFA, see][]{sanchez2012,
Husemann2013, GarciaBenito2015}.  These data have enabled studies of
the radial and azimuthal distributions of different characteristics in
galaxies such as their stellar populations
\citep{SanchezBlazquez2014,MartinNavarro2015}, star formation rate and
history \citep{Perez2013,GonzalezDelgado2014,GonzalezDelgado2016},
motions of the ionized gas
\citep{BarreraBallesteros2015,GarciaLorenzo2015,Holmes2015}, and
oxygen abundances
\citep{Sanchez2012b,Pilyugin2014,Sanchez2014,Zinchenko2016,SanchezMenguiano2016a}. 	

In our current study we will construct maps of the velocity field of
the ionized gas, of oxygen abundances, and of the extinction (Balmer
decrement) within the optical radius, $R_{25}$, in the discs of four
galaxies with two-dimensional spectroscopy from the CALIFA survey. We
will search for the manifestation of spiral arms using the method of
Fourier analysis of the azimuthal distribution of the tracers
(non-circular motions, oxygen abundance, and extinction) in thin ring
zones at different galactocentric distances in the plane of the
galaxy.  The decomposition of the two-dimensional maps of different
tracers through Fourier analysis allows us to quantify structures
in a range of scales from $\sim 0.2 R_{25}$ to $\sim 0.7 R_{25}$.  The
Fourier analysis provides data on the multiplicity (simultaneous
existence of several modes), geometrical form, and radial extent of
the spiral arms.  Such an analysis was used before to determine the
spiral arms from the positions of H\,{\sc ii} regions
\citep{considere1982}, the light distribution at a given wavelength
\citep{considere1988}, and velocity fields of ionized gas
\citep*{sakhibov1987, sakhibov1989, sakhibov1990, sakhibov2004,
canzian1993, fridman2001a, fridman2001b}.  Our aim is to determine the
parameters of the spiral arms based on different tracers, and to study
their morphological relations.

The paper is structured as follows. The data are discussed in
Section~\ref{sect:data}.  In Section~\ref{sect:kinematic} we describe
the method of the Fourier analysis of a two-dimensional velocity field
and the results obtained for our target galaxies.  The determination
of the spiral arms from the oxygen abundance and the extinction maps
is described in Section~\ref{sect:Z_Av_tracers}.  The comparison
between spiral arms derived from different tracers is given in
Section~\ref{sect:discussion}.  Section~\ref{sect:conclusions}
summarizes our conclusions.

%++++++++++++++++++ Table1    Galaxies
\begin{table*}
\caption[]{\label{table:sample}
The galaxy sample.
}
\begin{center}
\begin{tabular}{cccccccccccc} \hline \hline
Galaxy   & Type &T-type& log$(M_*)^a$& $M_R^b$ &Inclination$^c$& PA$^c$ &$R_{25}^c$&$R_{25}^c$ & $V_{max}^d$  & $D$   \\
         &      &      &            & [mag]   & [degree]
&[degree]& [arcmin] & [kpc]  & [km s$^{-1}$]  & [Mpc]  \\
       1 & 2    & 3    & 4          & 5       & 6             & 7      & 8        & 9        & 10  & 11 \\
\hline
NGC~36   & SABb & 3.0  & 10.82      & -22.34  & 64            &  14    & 0.95     & 22.38 &253.8    & 81.0 \\
NGC~180  & Sc   & 4.6  & 10.66      & -22.27  & 46            & 165    & 1.09     & 22.39  &217  & 70.6   \\
NGC~6063 & Scd   & 5.9  &  9.94      & -20.54  & 56            & 156    & 0.75     & 10.19  & 141.6 & 46.7  \\
NGC~7653 & Sb   & 3.1  & 10.48      & -21.62  & 31            & 175    & 0.77     & 13.06  & 228  & 58.3  \\
\hline
\end{tabular}\\
\end{center}
\begin{flushleft}
$^a$ Stellar mass estimated by the CALIFA collaboration \citep{walcher2014} \\
$^b$ R-band absolute magnitude obtained by the CALIFA collaboration \citep{walcher2014} \\
$^c$ Values taken from \citet{Zinchenko2016} \\
$^d$ Maximum rotation velocity corrected for inclination ( http://leda.univ-lyon1.fr/search.html) 
\end{flushleft}
\end{table*}

\begin{figure}
%\vspace{1.0mm}
%\resizebox{1.00\hsize}{!}{\includegraphics[angle=000]{Fig01.eps}}
\resizebox{1.00\hsize}{!}{\includegraphics[angle=000]{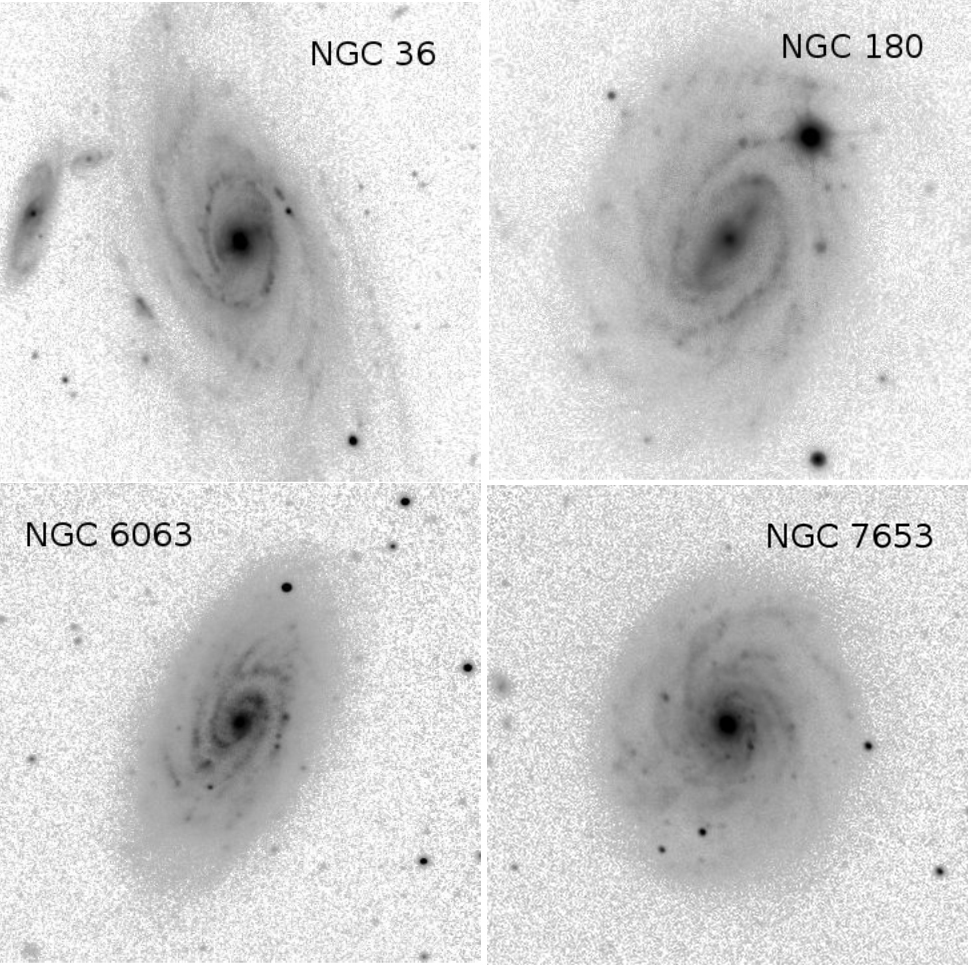}}
\caption{Images of the target galaxies are shown in the Sloan Digital Sky Survey (SDSS) $g$ filter ($ 4686\AA$).
The images are taken from the  NASA Extragalactic Database (NED).  
North is up and east is to the left.  
}
\label{figure:pictures}
\end{figure}

\begin{figure*}
%\begin{figure}
%\vspace{1.0mm}
%\resizebox{1.00\hsize}{!}{\includegraphics[angle=000]{Fig02.eps}}
\resizebox{0.90\hsize}{!}{\includegraphics[angle=000]{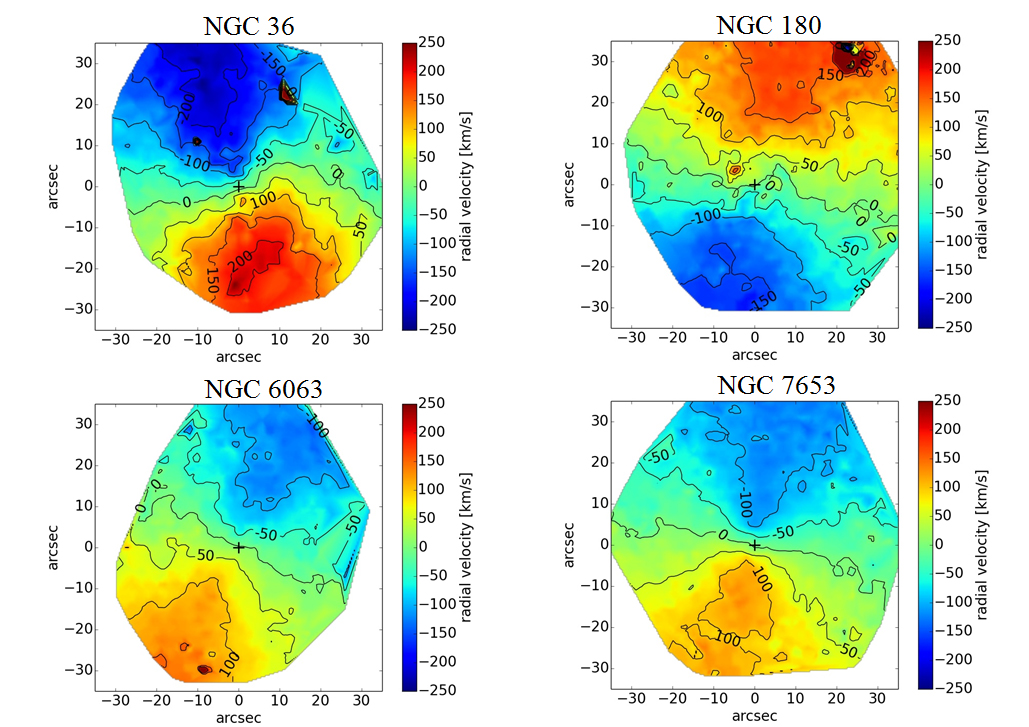}}
\caption{
The colour-coded maps of the observed velocity fields in the target
galaxies corrected for the systemic velocity of the respective galaxy
as a whole.  The lines show the contours of equal line-of-sight
velocities. The contours are separated by $50$~km~s$^{-1}$.  The
position of the centre of the galaxy is marked by a plus sign. }
\label{figure:map_2D_V}
%\end{figure}
\end{figure*}

\section{The Data}
\label{sect:data}
%======================

We selected four galaxies from the sample of CALIFA galaxies studied
in \citet{Zinchenko2016}.  We chose galaxies where the spaxels with
measured emission lines are homogeneously distributed over the image
of the galaxy.  Images of the target galaxies in the Sloan Digital Sky
Survey (SDSS) $g$ filter are shown in Fig.~\ref{figure:pictures}.
The images are taken from the  NASA Extragalactic Database 
(NED)\footnote{https://ned.ipac.caltech.edu/}.  
The processing of the observational data is reported in
\citet{Zinchenko2016}.  Therefore we just briefly describe the data
analysis steps and report additional measurements performed here.

We used publicly available spectra from the CALIFA survey data release
2 (DR2; \cite*{sanchez2012, GarciaBenito2015,walcher2014}), based on
observations with the PMAS/PPAK integral field spectrophotometer
mounted on the Calar Alto 3.5-m telescope in Spain.  We used the
low-resolution data cubes (setup V500) to determine the spatial
distribution of the physical parameters of the interstellar ionized
gas.  The integrated stellar spectra in all spaxels were fitted using
the public version of the {\sc STARLIGHT} code
\citep{CidFernandes2005,Mateus2006,Asari2007} adapted for execution in
the NorduGrid ARC\footnote{http://www.nordugrid.org/} environment of
the Ukrainian National Grid.  We used a set of 45 synthetic simple
stellar population (SSP) spectra with metallicities $Z = 0.004$, 0.02,
and 0.05, and 15 ages from 1~Myr up to 13~Gyr from the evolutionary
synthesis models of \citet{BC03}, and the reddening law of
\citet[]{CCM} with $R_V = 3.1$.  The nebular emission spectrum in each
spaxel was obtained by subtracting the stellar population synthesis
spectrum from the observed one.  The profiles of H$\beta$, H$\alpha$,
[OIII]$\lambda\lambda$4959,5007, [NII]$\lambda\lambda$6548,6584, and
[SII]$\lambda\lambda$6717,6731 lines were fitted by Gaussians.

The line-of-sight velocity of the interstellar gas was determined from
the wavelength of the centre of the H$\alpha$ line profile. The
observed velocity fields of the target galaxies corrected for the
systemic velocity of each galaxy, $V_{sys}$, are shown in
Fig.~\ref{figure:map_2D_V}.  We list the inferred values of $V_{sys}$
in Table~\ref{table:result_kinematic}.

The measured line fluxes were corrected for interstellar reddening
using the theoretical H$\alpha$ to H$\beta$ ratio from
\cite{osterbrock1989}, assuming case B recombination, an electron
temperature of 10,000 K, and the analytical approximation to the
Whitford interstellar reddening law by \citet{izotov1994}.  If the
measured value of the ratio H$\alpha$/H$\beta$ is lower than the
theoretical one (2.86) then the reddening is adopted to be zero. 

We used the oxygen abundances determined in \cite{Zinchenko2016}
through a strong-line method \citep[the $C$ method,][]{Pilyugin2012}.
It should be noted that the $C$ method produces oxygen abundances
compatible to the metallicity scale defined by H\,{\sc ii} regions
with $T_{e}$-based abundances.  

 For the further analysis of the velocity field and the extinction
map we selected spaxels with a ratio of the flux to the flux error
$\epsilon \ge 8$ for the H$\alpha$ emission line. For the oxygen
abundance map, only spaxels with $\epsilon \ge 3$ for the each of
the H$\beta$, H$\alpha$, [OIII]$\lambda$5007, [NII]$\lambda$6584 lines
were selected for further analysis.

The properties of our target galaxies are given in
Table~\ref{table:sample}.  The galaxy name is reported in the first
column.  The morphological type and the morphological type code $T$ of
the galaxy taken from the LEDA data
base\footnote{http://leda.univ-lyon1.fr/}  \citep{paturel2003} are
reported in Columns 2 and 3. The stellar mass of the galaxy and
its absolute magnitude are listed in Columns 4 and 5.  The galaxy
inclination and position angle of the major axis obtained from the
analysis of the photometric map are reported in Columns 6 and 7.
The isophotal radii $R_{25}$ in arcmin and in kpc are given in Columns
8 and 9, respectively.  The adopted distance is listed in Column 10.

Our subsequent study is based on the maps of velocity, oxygen
abundance, and extinction $A_{V}$ (Balmer decrement) of the four
galaxies.

\begin{figure}
%\vspace{1.0mm}
%\resizebox{1.00\hsize}{!}{\includegraphics[angle=000]{Fig03.eps}}
\resizebox{1.00\hsize}{!}{\includegraphics[angle=000]{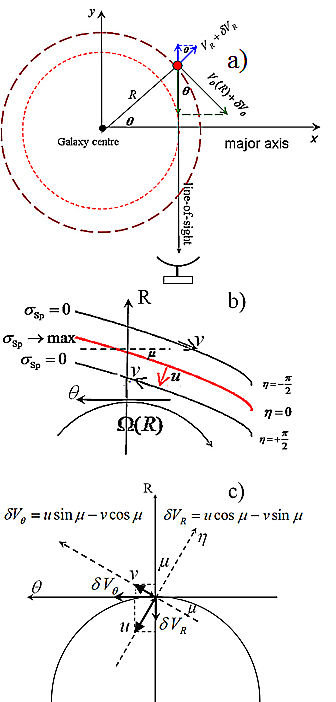}}
\caption{ 
$a)$  Contributions of the different components of the gas motion to 
the line-of-sight velocity.
$b)$ The velocity field of a spiral density wave. $\sigma_{Sp}$
denotes the density perturbation of the galactic disc.  The pitch
angle of the spiral arm, $\mu$, is defined as the angle between
tangents to the circle and to the spiral arm.  The lines of equal
spiral arm phase, $\eta = const$, are lines of equal density and
velocity perturbation.  The component of the streaming velocity
perpendicular to the spiral arm, $u$, has its extreme value at $\eta
=0$.  The tangential streaming velocity, $v$, reaches its extreme
values for $\eta = \pm \pi/2$, i.e., at the outer and inner spiral arm
borders (perturbed density $\sigma_{Sp} = 0$). 
$c)$ The radial (along the direction of increasing galactocentric
distance) and tangential (parallel to the direction of rotation) components of
the velocity perturbations ($\delta V_R$, $\delta V_{\theta}$) are
related to the streaming velocity $(u, v)$ by 
Eq.~\ref{equation:dV_theta_dV_rad} and depend on the pitch angle
$\mu$.  
}
\label{figure:method}
\end{figure}

\section{The kinematic spiral arms}
\label{sect:kinematic}

\subsection{Fourier coefficients from the velocity field }
\label{sect:velocity_field}

The Fourier transformation method is used to decompose a two-dimensional
velocity field into a hierarchy of structures on different scales.
This method, applied to a number of narrow concentric annuli in radial
direction of the galaxy, allows one to extract the field of
perturbations in both the angular and radial components of the
velocity and to investigate logarithmic and other types of spiral
arms. It should be noted that there is no assumption of the shape of
the spiral arms with this method.  In the current Section this method
is applied to the two-dimensional velocity fields to investigate the
spiral structure of the velocity perturbations arising due to spiral
density waves.

The line-of-sight velocity field of the gas in galactic discs exhibits
deviations from a purely circular motion, which are associated with
the spiral arms of the galaxy \citep{visser1980}. These deviations
could be due to the influence of a spiral density wave on the
circular rotation of the gas in the disc \citep{lin1969}.  The
tangential and radial components of perturbed velocities ($\delta
V_{\theta}$, $\delta V_R$) caused by a spiral density wave depend on
the amplitude of the wave, the position of the corotation and Lindblad
resonances, and the mass distribution within the galaxy.  Here $R$ is
the deprojected galactocentric distance in the plane of the galaxy and
$\theta$ is the polar angle, measured from the major axis of the
galaxy. 
 There can also be deviations from pure circular motions of other
types, e.g., a radially symmetric outflow or inflow toward the center 
of the galaxy, caused by bars, warps, lopsidedness, or a galactic fountain
\citep[e.g.,][]{bosma1978, sakhibov1987, sakhibov1989, sakhibov1990,  
sakhibov2004, fraternali2001, schinnerer2000}. 
The description of different approaches to the modelling of non-circular 
motions in a two-dimensional velocity map is given in \citet{spekkens2007}, 
where the non-circular motions with rather high velocities due to 
a bar-like distortion of an axisymmetric potential are considered. 
The objective of the present paper is to examine the non-circular motions 
caused by density wave perturbations. It is expected that the 
radial velocities caused by a bar can exceed the non-circular velocities 
caused by density wave perturbations in the bar region. 
We will investigate the non-circular motions in the disc outside the bars
where radial velocities caused by the spiral perturbation are expected to make 
a dominant contribution to the non-circular motions.

The contributions of the perturbed tangential ($\delta V_{\theta}$)
and radial ($\delta V_R$) components of the velocity to the observed
line-of-sight velocity are given by the equation  (see
Fig.~\ref{figure:method}a) 
\begin{equation}
V_{sp}= \Bigl(\delta V_{\theta}(R,\theta) \cdot \cos \theta +\delta V_R (R,\theta) \cdot \sin \theta \Bigl ) \cdot \sin i
\label{equation:V_Sp}
\end{equation}
Thus, the line-of-sight velocity  $V^{obs}_r(R,\theta)$ observed at a
given point of the disc involves the velocity of the galaxy as a whole,
$V_{gal}$, the rotation velocity, $V_{\theta}(R)$,  a pure radial
inflow or outflow $V_R$, and the velocities $\delta V_{\theta}$ and $\delta V_R$ caused by the density wave
perturbation (see Fig.~\ref{figure:method}a). 

\begin{equation}
%\frac{V^{obs}_r(R,\theta) }{\sin i}= V_{gal} +\Bigl( V_{\theta}(R)+\delta V_{\theta}(R,\theta) \Bigl ) \cdot \cos \theta +\delta V_R (R,\theta) \cdot %\sin \theta 
\frac{V^{obs}_r(R,\theta) }{\sin i}= V_{gal} +\Bigl( V_{\theta}(R)+\delta V_{\theta}(R,\theta) \Bigl ) \cdot \cos \theta 
+\Bigl( V_R+\delta V_R(R,\theta) \Bigl ) \cdot \sin \theta
\label{equation:v_light_sight}
\end{equation}
where $i$ is the inclination angle of the galaxy.

Our aim is to extract the contributions caused by the spiral density
wave, $\delta V_{\theta}$ and $\delta V_R$, to the observed velocity
$V^{obs}_r(R,\theta)$.  If the gravitational potential of a disc
galaxy is perturbed by a rotating spiral wave, the gas in the disc
reacts by streaming motions that are also of spiral shape. The
response of the gas then is
\begin{eqnarray}
\label{equation:u_eta_u_xi}
u=\hat u \cos \eta \\
v= -\hat v \sin\eta \nonumber
\end{eqnarray}
where $\eta$ is the phase of the perturbation by a spiral wave (which varies
across the spiral arm, see Fig.~\ref{figure:method}b)
\begin{eqnarray}
\label{equation:eta_xi}
 \eta=-m\theta +mh\cot \mu \ln{(R/R_0)}  
\end{eqnarray}
where $h = +1$ and $h = - 1$  for an S- and Z-spiral arm,
respectively, $\mu$ is the pitch angle (a measurement of the tightness of
the winding of spiral arms), and $m$ is the number of the arms.  
The parameter h determines the direction of winding of the spiral arm 
in the sky: S-shaped or Z-shaped.
Note that a right-handed coordinate system is used.
The values $\hat u$ and $\hat v$ are amplitudes (extreme values) of
orthogonal (across the spiral arm) and tangential streaming
velocities.  The orthogonal streaming velocity has its extreme value
at the centre of the spiral arm for $\eta=0$ and the tangential
streaming velocity has its extreme values at the outer and inner
spiral arm borders  for $\eta=\pm \pi /2$ (Fig.~\ref{figure:method}b).

The orthogonal and tangential streaming velocities make a contribution
to the observed line-of-sight velocity $V^{obs}_r(R,\theta)$ in the
form of the velocity perturbations ($\delta V_R$, $\delta
V_{\theta}$).  The radial (along the direction of increasing
galactocentric distance) and tangential (parallel to the direction of 
rotation) components of the velocity perturbations ($\delta V_R$,
$\delta V_{\theta}$) are related to the streaming velocity $(u, v)$ by
Eq.~\ref{equation:dV_theta_dV_rad} and depend on the geometrical
form (pitch angle $\mu$) of the spiral arm (see
Fig.~\ref{figure:method}c) 
\begin{eqnarray}
\label{equation:dV_theta_dV_rad}
 \delta V_{\theta}= h(u \sin \mu - v\cos \mu )  \\
 \delta V_R= u\cos \mu - v\sin \mu  \nonumber
\end{eqnarray}

Using Eq.~(\ref{equation:u_eta_u_xi}) to
Eq.~(\ref{equation:dV_theta_dV_rad}) and grouping the sine and cosine
terms of identical quantities of $\theta$, the perturbed motion
$V_{sp}$ (Eq.~(\ref{equation:V_Sp})) caused by the mode $m$ can be
written in the form
\begin{eqnarray}
\label{equation:V_Sp2}
V_{sp}=\frac{1}{2}(\hat u- \hat v )\cdot \cos\Bigl((m-1)\theta \Bigr) \cdot \sin\Bigl(m\cot \mu \ln{(R/R_0) + \mu\Bigr)} \nonumber\\
            -\frac{1}{2}(\hat u- \hat v )\cdot \sin\Bigl((m-1)\theta \Bigr) \cdot \cos\Bigl(m\cot \mu \ln{(R/R_0) + \mu\Bigr)} \\
          -\frac{1}{2}(\hat u+ \hat v )\cdot \cos\Bigl((m+1)\theta \Bigr) \cdot \sin\Bigl(m\cot \mu \ln{(R/R_0) - \mu\Bigr)} \nonumber\\
         +\frac{1}{2}(\hat u+ \hat v )\cdot \sin\Bigl((m+1)\theta \Bigr) \cdot \cos\Bigl(m\cot \mu \ln{(R/R_0) - \mu\Bigr)} \nonumber
\end{eqnarray}

Eq.~(\ref{equation:V_Sp2}) corresponds the case of $h=+1$, because all
our target galaxies are S-spirals.  The coefficients of the sines and
cosines of the polar angles $(m-1)\theta$ and $(m + 1)\theta$
determine the contribution of the velocity perturbations to the
observed line-of-sight velocity.  Inspection of
Eq.~(\ref{equation:V_Sp2}) shows that the second mode of the
spiral-density wave ($m = 2$) -- a two-armed spiral -- contributes to
the values multiplied by the sines and cosines of the polar angles
$\theta$ and $3\theta$ while the first mode ($m = 1$) -- a one-armed
spiral -- contributes to the values multiplied by the sines and
cosines of the polar angles $\theta =0 $ and $2\theta$.

Eq.~(\ref{equation:V_Sp2}) and Eq.~(\ref{equation:v_light_sight}) result in 
\begin{eqnarray}
\label{equation:V_line-of-sight_2}
\frac{ V^{obs}_r(R,\theta)}{\sin i} = a_0+a_1 \cos(\theta)+ b_1 \sin (\theta)+\nonumber \\
                                                     + a_2 \cos (2\theta)+ b_2 \sin (2\theta) + a_3 \cos (3\theta)+ b_3 \sin (3\theta) 
\end{eqnarray}
where the  coefficients $a_0,a_1,b_1,a_2,b_2, a_3$, and $b_3$ are 
\begin{eqnarray}
\label{equation:Fourieir_Coef}
a_0=V_{gal} +\frac{1}{2}(\hat u_1- \hat v_1 )\cdot \sin\Bigl(\cot \mu_1 \ln{(R/R_{01}) + \mu_1\Bigr)}   \nonumber\\
a_1=V_{\theta}(R)+ \frac{1}{2}(\hat u_2 - \hat v_2) \cdot \sin\Bigl(2\cot \mu_2 \ln{(R/R_{02}) + \mu_2\Bigr)} \nonumber \\
b_1=V_R - \frac{1}{2}(\hat u_2- \hat v_2) \cdot \cos\Bigl(2\cot \mu_2 \ln{(R/R_{02}) + \mu_2\Bigr)} \\
%\nonumber\\
a_2= -\frac{1}{2}(\hat u_1+\hat v_1)\cdot \sin\Bigl(\cot \mu_1 \ln{(R/R_{01}) - \mu_1 \Bigr)} \nonumber \\
b_2= \frac{1}{2} (\hat u_1+\hat v_1) \cdot \cos\Bigl(\cot \mu_1 \ln{(R/R_{01}) - \mu_1 \Bigr)} \nonumber\\
a_3=-\frac{1}{2}(\hat u_2+\hat v_2)\cdot \sin\Bigl(2\cot \mu_2 \ln{(R/R_{02}) - \mu_2 \Bigr)} \nonumber \\
b_3= \frac{1}{2}(\hat u_2+\hat v_2) \cdot \cos\Bigl(2\cot \mu_2 \ln{(R/R_{02}) - \mu_2 \Bigr)} \nonumber
\end{eqnarray}
Hereafter the quantities $\hat u_1$, $\hat v_1$, $\mu_1$, and $R_{01}$
mean the velocity perturbation amplitudes, the pitch angle, and
scaling factor for the first mode ($m=1$), and $\hat u_2$, $\hat v_2$,
$\mu_2$, and $R_{02}$ stand for the velocity perturbation amplitudes,
the pitch angle, and the scaling factor for the second mode ($m=2$) of
a spiral density wave, respectively.

The systemic velocity $V_{gal}$ of the galaxy (not corrected for inclination) contributes in the
zeroth Fourier harmonic and mixes with the velocity perturbation from the first mode $m=1$ of the spiral density wave.  

The circular rotation $V_{\theta}(R)$ contributes in the first Fourier
harmonics ($a_1$) and mixes  with the velocity perturbation from the second mode $m=2$ of the spiral density wave.

The first mode of
the spiral wave ($m = 1$) -- a one-armed asymmetric structure --
contributes to the zeroth and second Fourier harmonics of the
azimuthal distribution of the observed line-of-sight velocities.  The
second mode of the spiral-density wave ($m = 2$) -- a two-armed spiral
-- contributes also to the first and third Fourier harmonics of the
azimuthal distribution of line-of-sight velocities at a given
galactocentric distance, $R$.  Since the zeroth Fourier coefficient,
$a_0$, may include a contribution not only from the systemic velocity
of the galaxy $V_{gal}$ but also from the first mode of the spiral
density wave, the zeroth Fourier coefficient must change in radial 
direction along the galactic disc.

Thus, if the velocity of the galaxy, $V_{gal}$, and the
radial change of the Fourier coefficients, $a_1(R)$, are determined,
the pure rotation curve of the galaxy, $V_{\theta}(R)$, can be
measured.  The radial variation of the coefficients $a_2(R), b_2(R),
a_3(R)$, and $b_3(R)$ defines the amplitude of the velocity
perturbations caused by the spiral arms as well as the geometric form
of the arms.  Thus, the problem of the analysis of the residual
velocities caused by a spiral density wave can be reduced to the
determination of the first three harmonics of the Fourier series of
the observed azimuthal distribution of the line-of-sight velocities
for the thin ring zones.

There is also an impact of the bar in the inner part of a barred galaxy.
Accounting for the impact of the bar, according to the bisymmetric model by \citet{spekkens2007}, 
changes Eq.~\ref{equation:v_light_sight}  for the observed line-of-sight velocity:
\begin{eqnarray}
\label{equation:V_line-of-sight+Bar}
\frac{V^{obs}_r(R,\theta) }{\sin i}= V_{gal} +\Bigl( V_{\theta}(R)+\delta V_{\theta}(R,\theta) +V_{2,t} \Bigl ) \cdot \cos \theta + \nonumber \\
						+\Bigl( V_R+\delta V_R(R,\theta) + V_{2,r} \Bigl ) \cdot \sin \theta
\end{eqnarray}
where $V_{2,t} = \hat V_{2,t}\cos(2 \theta_b)$ and $V_{2,r}=\hat V_{2,r}\sin(2 \theta_b)$ 
are the tangential and radial components of the non-circular flow
caused by the bar.  These components vary with the angle to the bar 
axis $\theta_b= \theta - \phi_b$ as the cosine and sine of $2\theta_b$, respectively.
The major axis of the bar lies at an angle $\phi_b$ to the projected
major axis of the galaxy (see Fig.~1 in \citet{spekkens2007}).
The maximal amplitudes of the tangential and radial components of the
non-circular flow caused by bar, 
$\hat V_{2,t}(R)$ and $\hat V_{2,r(R)}$, respectively, are both
functions of the radius. 
Eq.~\ref{equation:Fourieir_Coef} can be rewritten as:
\begin{eqnarray}
\label{equation:Fourieir_Coef+Bar}
a_0=V_{gal} +\frac{1}{2}(\hat u_1- \hat v_1 )\cdot \sin\Bigl(\cot \mu_1 \ln{(R/R_{01}) + \mu_1\Bigr)}   \nonumber\\
a_1=V_{\theta}(R)+ \frac{1}{2}(\hat u_2 - \hat v_2) \cdot \sin\Bigl(2\theta_{Sp2} + \mu_2\Bigr)- \frac{1}{2}(\hat V_{2,t} - \hat V_{2,r}) \cdot \sin\Bigl(2 \phi_b\Bigr)  \nonumber \\
b_1=V_R - \frac{1}{2}(\hat u_2- \hat v_2) \cdot \cos\Bigl(2\theta_{Sp2} + \mu_2\Bigr)+ \frac{1}{2}(\hat V_{2,t} - \hat V_{2,r}) \cdot \cos\Bigl(2 \phi_b\Bigr) \\
%\nonumber\\
a_2= -\frac{1}{2}(\hat u_1+\hat v_1)\cdot \sin\Bigl(\cot \mu_1 \ln{(R/R_{01}) - \mu_1 \Bigr)} \nonumber \\
b_2= \frac{1}{2} (\hat u_1+\hat v_1) \cdot \cos\Bigl(\cot \mu_1 \ln{(R/R_{01}) - \mu_1 \Bigr)} \nonumber\\
a_3=-\frac{1}{2}(\hat u_2+\hat v_2)\cdot \sin\Bigl(2\theta_{Sp2} - \mu_2 \Bigr) -  \frac{1}{2}(\hat V_{2,t} + \hat V_{2,r}) \cdot \sin\Bigl(2 \phi_b\Bigr)  \nonumber \\
b_3= \frac{1}{2}(\hat u_2+\hat v_2) \cdot \cos\Bigl(2\theta_{Sp2} - \mu_2 \Bigr) + \frac{1}{2}(\hat V_{2,t} + \hat V_{2,r}) \cdot \cos\Bigl(2 \phi_b\Bigr)  \nonumber
\end{eqnarray}
where  $\theta_{Sp2}=\cot\mu_2 \ln{(R/R_{02})}$ belongs to the
bisymmetrical (mode $m=2$) of the spiral perturbations.
Eq.~\ref{equation:Fourieir_Coef+Bar} shows that the spiral term at the coefficients $a_1, b_1, a_3, b_3$ is 
modulated by the periodic sine or cosine functions. 
The amplitudes of the tangential and radial components of the
non-circular flow caused by the bar, 
$\hat V_{2,t}(R)$ and $\hat V_{2,r(R)}$, respectively, are both
functions of the radius. 
Since the elliptical bar potential decreases monotonously with radius
$R$ the bar term in Eq.~\ref{equation:Fourieir_Coef+Bar} also
decreases monotonously.
Eq.~\ref{equation:v_light_sight} and Eq.~\ref{equation:Fourieir_Coef}
correspond to the radial model by \citet{spekkens2007} 
with additional terms of spiral perturbations. 
Eq.~\ref{equation:V_line-of-sight+Bar} and Eq.~\ref{equation:Fourieir_Coef+Bar} 
correspond to the bisymmetric model by the same authors with additional terms of spiral perturbations.
We refer to the model described by Eq.~\ref{equation:v_light_sight} and Eq.~\ref{equation:Fourieir_Coef} 
as the radial model with spiral perturbations and to the model described by Eq.~\ref{equation:V_line-of-sight+Bar} and 
Eq.~\ref{equation:Fourieir_Coef+Bar} as the bisymmetric model with spiral perturbations.

In the outer part of the galaxy, the impact of the elliptical bar
potential decreases quickly with radius $R$. 
A bar term would add to the radial symmetric (outflow/inflow) motion and the impact of the spiral term, or vanish.
This means that beyond the bar Eq.~\ref{equation:V_line-of-sight+Bar} and Eq.~\ref{equation:Fourieir_Coef+Bar} will 
coincide with Eq.~\ref{equation:v_light_sight} and Eq.~\ref{equation:Fourieir_Coef}.
Since our main goal is a study of the impact of spirals on the spatial
distribution of tracers, 
we adopt the radial model with spiral perturbations (Eq.~\ref{equation:v_light_sight} and Eq.~\ref{equation:Fourieir_Coef}) outside the bar.
The pure radial outflow or inflow velocities contribute to the first Fourier harmonic ($b_1$) while   
the pitch angle $\mu_2$ and characteristic radius $R_{02}$ will be determined from the third Fourier harmonic. 
The radial disturbance by the bar has no impact on the definition of
the geometrical parameters of the spiral arms.

\begin{figure}
%\vspace{1.7mm}
%\resizebox{1.00\hsize}{!}{\includegraphics[angle=000]{Fig04.eps}}
\resizebox{1.00\hsize}{!}{\includegraphics[angle=000]{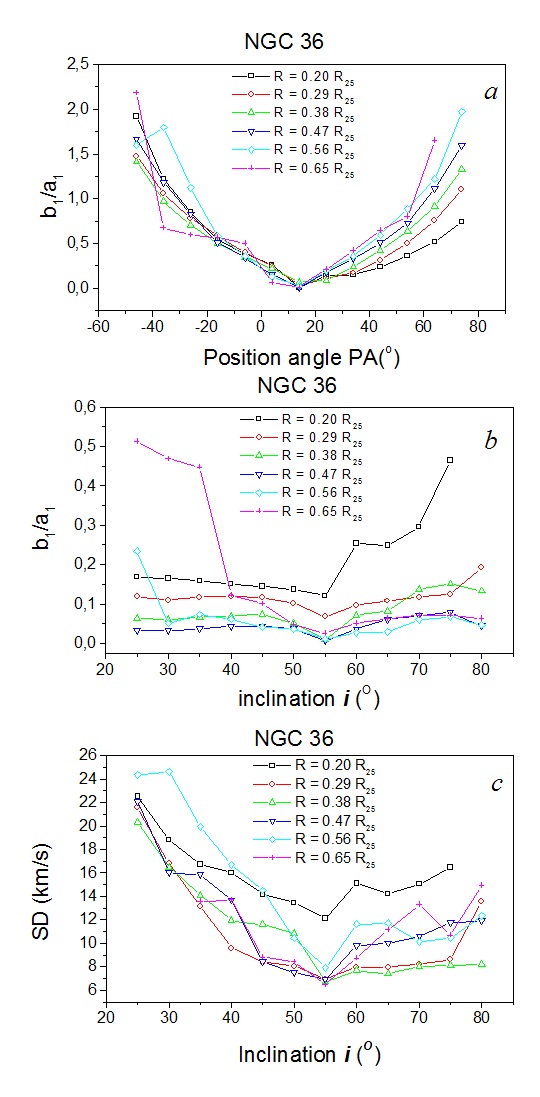}}
\caption{
The change of the ratio of the Fourier coefficients $b_1/a_1$ in
dependence on the proposed values of position angle $PA$ (a)
 and inclination angle $i$ (b) in NGC~36.
c)The change of the standard deviation $SD$ of the observed
line-of-sight velocities from the predicted ones in dependence on the
proposed values of the inclination angle $i$ in NGC~36.
}
\label{figure:PA}
\end{figure}

\subsection{Determination of the galaxy inclination and the position 
angle of the major axis}
\label{sect:PA_Inclination}

Here we estimate the galaxy inclination and the position angle of the major axis from the analysis of the observed velocity field and 
compare them with the values for the same quantities obtained from the analysis of the surface brightness distribution \citep{Zinchenko2016}.
One may expect that the rotation makes a dominant contribution to the observed line-of-sight velocities. 
This means that the ratio of the Fourier coefficients $b_1/a_1$ is minimized
if the galaxy inclination and position angle of the major axis are correct.
It is expected that the velocities of the radially symmetric motions $V_R$ are
within 10 -- 20 km~s$^{-1}$, the velocities of the spiral perturbations $\delta V_R$  
within  5 -- 10 km~s$^{-1}$ \citep{rohlfs1977}, and the rotation velocities $V_{\theta}$ within
150 -- 300 km~s$^{-1}$. Then the expected value of the ratio $b_1/a_1$ does not exceed 
$10\% - 15\%$ if the values of $i$ and $PA$ are correct. 
The radial velocities can reach unrealistically large values (which
may be comparable to the 
rotation velocities or even exceed them) if the values of $i$ and $PA$ 
are not correct. 
Thus, $PA$ and $i$ are estimated from the requirement that the ratio of 
the Fourier coefficients $b_1/a_1$ is minimized for the correct values of $i$ and $PA$.

We first consider a set of values of galaxy inclinations and
position angles of the major axis with large steps, 
$\Delta i =5\degr$ and $\Delta P =10\degr$. 
A set of deprojected coordinates of individual pixels in the target galaxy 
for each set of the values of $i$ and $PA$ is obtained.  Then a set of the 
coefficients $a_0, a_1,b_1, a_2, b_2, a_3,$ and $b_3$ of
Eq.~(\ref{equation:V_line-of-sight_2}) is determined for each set of the deprojected coordinates 
of the individual pixels by fitting to the line-of-sight velocities. 
The disc is divided into annuli of width $\Delta R$ = 1 kpc in the plane of the 
target galaxy (see \ref{sect:spiral_arms_from_V_map}).
The coefficients $a_0, a_1,b_1, a_2, b_2, a_3,$ and $b_3$ are assumed to be 
equal for all the pixels within a given annulus.
Assuming that the rotation makes a dominant contribution to the observed 
line-of-sight velocities  we choose the values of $i$ and $PA$ 
that minimize the ratio of the Fourier coefficients $b_1/a_1$ 
as an approximation of the position angle of the major axis and the inclination 
angle of the galaxy.
Fig.~\ref{figure:PA}(a,b) shows the Fourier coefficient ratio $b_1/a_1$
as a function of $PA$ and $i$ for the galaxy NGC~36.
To avoid crowding in the diagrams, the data are presented for the interval from 4.5 kpc to 14.5 kpc 
(from $0.20R_{25}$ to $0..65R_{25}$) with a step size of 2 kpc. Each
curve corresponds to a single annulus. 
The step size in $PA$ is $\Delta_{PA}=10\degr$ and the step size in $i$ is $\Delta_i=5\degr$.
Each ring zone shows a different $b_1/a_1$ value, but every curve has its minimum 
at the same $PA=14\degr$ and at the same $i=55\degr$.
In some zones  the minimum of the curve in the $b_1/a_1$ vs.\ $i$ diagram 
is not so prominent as in the $b_1/a_1$ vs.\ $PA$ diagram. Therefore,
we consider an additional diagram.
Fig.~\ref{figure:PA}(c) shows the standard deviation {\it SD} of the observed 
line-of-sight velocity in an annulus compared to the one predicted by Eq.~\ref{equation:V_line-of-sight_2} 
as a function of inclination angle $i$ for NGC~36.
The minimum value of the {\it SD} occurs at the same inclination
$i=55\degr$ as in the previous  $b_1/a_1$ vs.\ $i$ diagram.
Every annulus shows different {\it SD} values, but almost all curves
have a similar minimum of {\it SD} $\approx$ (6--8) km~s$^{-1}$ at the same $i=55\degr$.

Next, we use a set of $PA$ and $i$ values with smaller step sizes of
$2\degr$ to find more accurate values of the $PA$ and $i$ for the
target galaxies. The obtained values of $PA$ and $i$ are listed in
Table~\ref{table:result_kinematic} (Columns 2 and 3, respectively) and
are used below to find the deprojected coordinates of the spaxels with
measured oxygen abundance and Balmer decrement. 

Our obtained values of $PA$ and $i$ are generally in agreement with
the photometry-based values of the $PA$ and $i$ reported in
Table~\ref{table:sample}.
The inclination of the galaxy NGC~36 determined here from 
the kinematics ($i=55\degr$) agrees with axis ratio $b/a= 0.56$ 
reported in the Uppsala General Catalogue of Galaxies  \citep{nilson1973} 
and is close to the $b/a= 0.62$ given in the Third Reference Catalogue 
of Bright Galaxies  \citep[RC3][]{Vaucouleurs1991}

 We only find a significant discrepancy between the values of the $PA$ determined by 
the different methods for the galaxy NGC~7653: $PA = 140\degr$
listed in 
the NED data base 
based on the angular diameters measured in $K_s$ (2MASS ``total'') passband,
$PA = 172.5\degr$ given in the LEDA data base,
$PA=175\degr$ based on the photometry map of \citet{Zinchenko2016}, 
and $PA=165\degr$ obtained here for the velocity map.
It should be noted that NGC~7653 is a nearly face-on galaxy.

\begin{figure}
%\resizebox{1.00\hsize}{!}{\includegraphics[angle=000]{Fig05.eps}}
\resizebox{1.00\hsize}{!}{\includegraphics[angle=000]{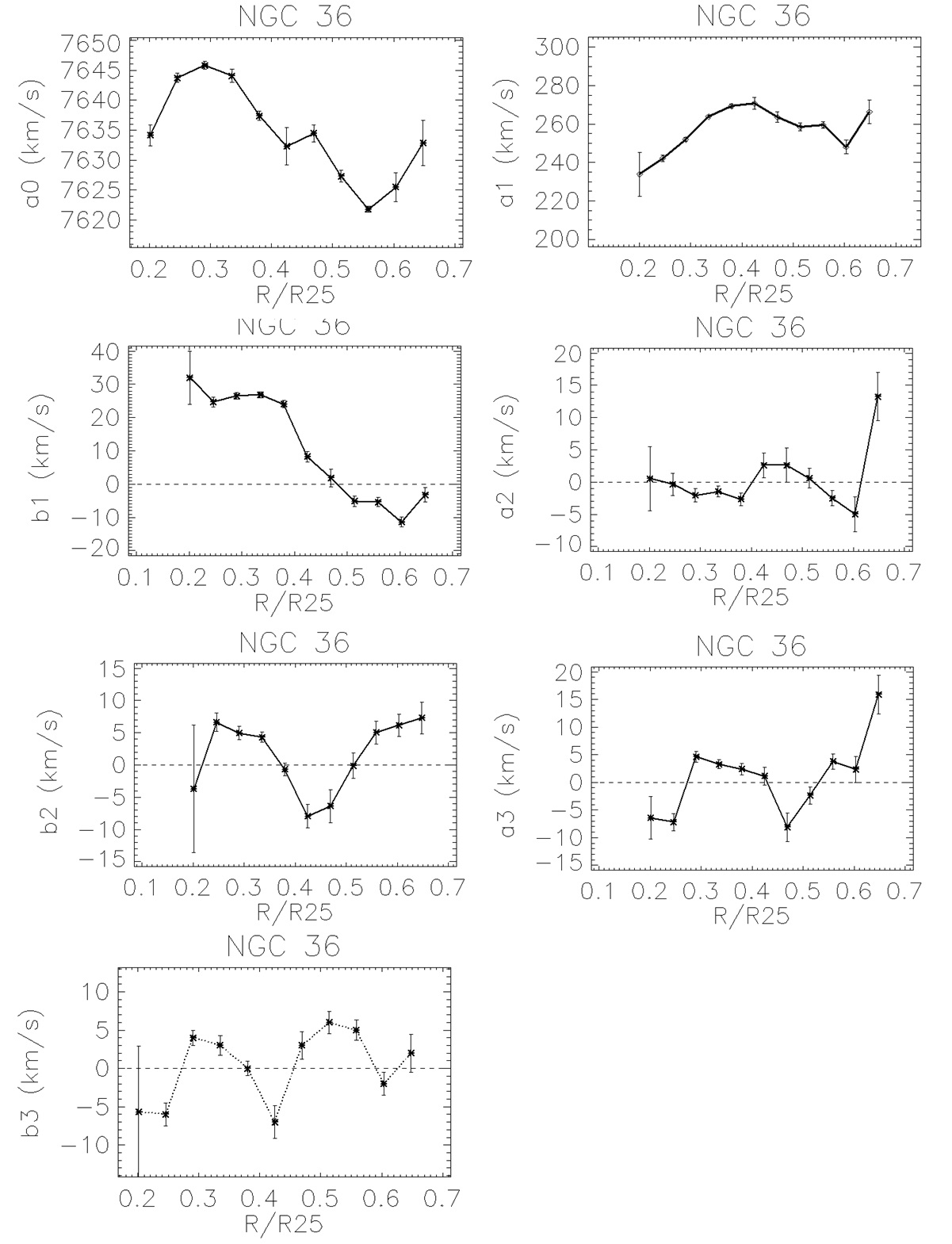}}
\caption{
The radial change of the Fourier coefficients 
 $a_0$,  $a_1$, $b_1$,  $b_2$,  $a_3$, and $b_3$ in NGC~36.
}
\label{figure:FC_N36}
\end{figure}

\subsection{Determination of the systemic velocity of galaxies from the velocity map}
\label{sect:systemic velocity}
%====================================================================

The deprojected coordinates of each pixel in the target galaxy are
determined with the position angle of the major axis and the galaxy
inclination obtained in Subsection~\ref{sect:PA_Inclination}.  
The galactic disc is divided into several annuli of a width of $\Delta R = 1$~kpc.
For every annulus, we determine the seven unknown Fourier coefficients 
$a_0, a_1,b_1, a_2, b_2, a_3,$ and $b_3$ of Eq.~(\ref{equation:V_line-of-sight_2}) 
using a least-squares method. 
The step size of $\Delta R = 1$~kpc provides 50 to 200 data points in each annulus in the 
interval of galactocentric distances from $R \sim 0.2R_{25}$ to $R \sim R=0.7R_{25}$. 
Thus the number of data points (spaxels) in each annulus is large enough for a reliable 
determination of the coefficients of Eq.~(\ref{equation:V_line-of-sight_2}).  
The determinations of the coefficients $a_0, a_1,b_1, a_2, b_2, a_3,$ and $b_3$ in a
several annuli allows us to assess the radial change of these coefficients.

Fig.~\ref{figure:FC_N36} shows an example of the radial change of the coefficients 
$a_0, a_1,b_1, a_2, b_2, a_3,$ and $b_3$ in case of NGC~36.
The range of galactocentric distances where the coefficients were
determined covers a large part of the disc from approximately 
$R=0.2R_{25}$ to $R=0.7R_{25}$.  As was noted in
Subsection~\ref{sect:velocity_field} the Fourier coefficients contain
information on the systemic velocity of the galaxy, $V_{gal}$, the pure
circular motion, $V_{\theta}(R)$, the radially symmetric motion, $V_R$, the peculiar velocities caused by the
spiral arms, and the geometric shape of the arms.

First we estimate the systemic velocity $V_{sys}$ of the galaxy,
which is a projection of the velocity of the galaxy as a
whole along the line-of-sight.
As noted above (Section~\ref{sect:kinematic}, Eq.\~(\ref{equation:Fourieir_Coef})),
the zeroth Fourier coefficient $a_0$
includes contributions from the first mode of the spiral density wave
and from the space velocity of the galaxy. 

If the true value of the systemic velocity $V_{sys}$ is subtracted
from the line-of-sight velocity of each point (spaxel) 
then the coefficient $a_0$ expresses the impact of the spiral wave only
and changes along the radius as a sine function 
\begin{equation}
a_0 = \frac{1}{2}\Bigl(\hat u_1- \hat v_1 \Bigr) \cdot \sin\Bigl(\cot \mu_1 \ln{(R/R_{01}) + \mu_1\Bigr)}  .
\label{equation:first_mode}
\end{equation}
The radially averaged value of $a_0$ must be close to zero, and 
the absolute value of maximum positive and negative amplitudes must be equal to each other.
This condition can be used to estimate the true value of the systemic velocity $V_{sys}$. 
To do so, we constructed a set of diagrams of $a_0$ vs.\ $R$ for different values of the $V_{sys}$.
The value of the $V_{sys}$ for which the absolute values of the maximum positive 
and negative values of the $a_0$ are equal to each other
is adopted to be true. 
In this case, the absolute value of the $a_0$ corresponds to the amplitude of the
non-circular perturbations caused by the spiral density wave.

\begin{figure}
%\vspace{1.7mm}
%\resizebox{1.00\hsize}{!}{\includegraphics[angle=000]{Fig06.eps}}
\resizebox{1.00\hsize}{!}{\includegraphics[angle=000]{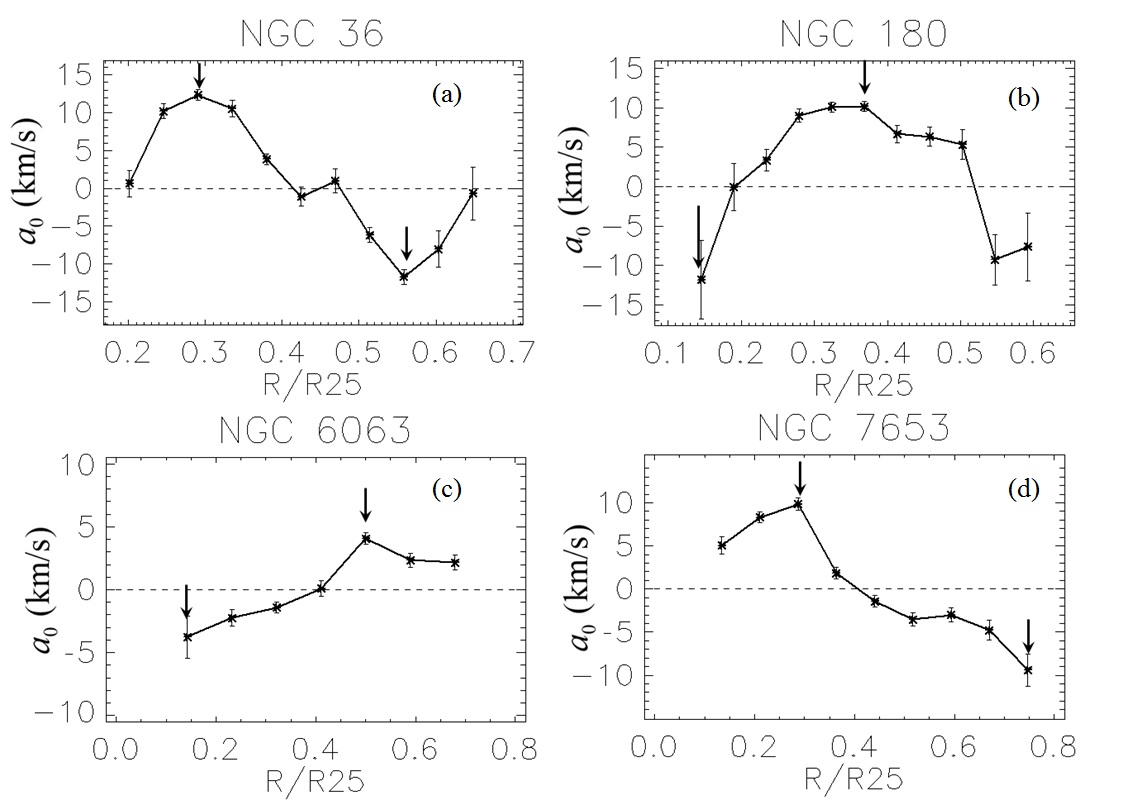}}
\caption{
The coefficient $a_0$,  computed for the true $V_{sys}$,  as a function of galactocentric distance
in our target galaxies.  The  $a_0(R)$ illustrates the radial change of the
 non-circular perturbations of the velocity caused by the first mode
($m=1$) of the spiral density wave (see
Eq.~\ref{equation:first_mode}).
The arrows show the positions of the maximum positive and negative amplitudes of perturbations, caused by the spiral density wave.
}
\label{figure:a0-Vsys}
\end{figure}

Fig.~\ref{figure:a0-Vsys} shows the value of  $a_0$ 
 obtained with the true $V_{sys}$
 as a function of galactocentric distance for each galaxy. 
We assume that Fig.~\ref{figure:a0-Vsys} shows the radial change of the 
non-circular perturbation caused by the first mode ($m=1$) of the spiral density 
wave in our target galaxies.
The bars in Fig.~\ref{figure:a0-Vsys} show the errors of the zeroth Fourier coefficients $a_0$, 
determined from the standard least-squares procedure. The large
number (50 -- 200) of the 
conditional equations (because of the large number of measurements) in
each annulus provides 
low uncertainties of $a_0$, i.e., the errors are significantly lower than 
the amplitude of the non-circular perturbations caused by the spiral density wave.
One can see from Fig.~\ref{figure:a0-Vsys} that the
amplitude of non-circular velocities is around 10~km~s$^{-1}$, which
agrees with the line-of-sight components of the density wave motions
predicted by the model (see Fig.\ 12 in \citet{rots1975}).  
This fact can be considered as additional evidence in favor 
of the validity of the value of $V_{sys}$.

Our estimates of the systemic line-of-sight velocities $V_{sys}$ agree 
with the values  of  other authors (listed in the NED database):
$6224\pm 12$ km~s$^{-1}$ for NGC~36 \citep{karachentsev1996},
$5394\pm 5$ km~s$^{-1}$ for NGC~180 \citep{Vaucouleurs1991},
$2944\pm 8$ km~s$^{-1}$ for NGC~6063 \citep{fixsen1996}, and
$4378\pm 20$ km~s$^{-1}$ for NGC~7653 \citep{{mould2000}}.

The obtained values of the systemic velocities of the target galaxies are
given in Column 4 of Table~\ref{table:result_kinematic}. 
  The uncertainty of the systemic velocity $V_{sys}$ is estimated as the sum of the error in the line-of-sight   velocity
 and the maximum amplitude of the first mode.

\begin{figure}
%\vspace{1.7mm}
%\resizebox{1.00\hsize}{!}{\includegraphics[angle=000]{Fig07.eps}}
\resizebox{1.00\hsize}{!}{\includegraphics[angle=000]{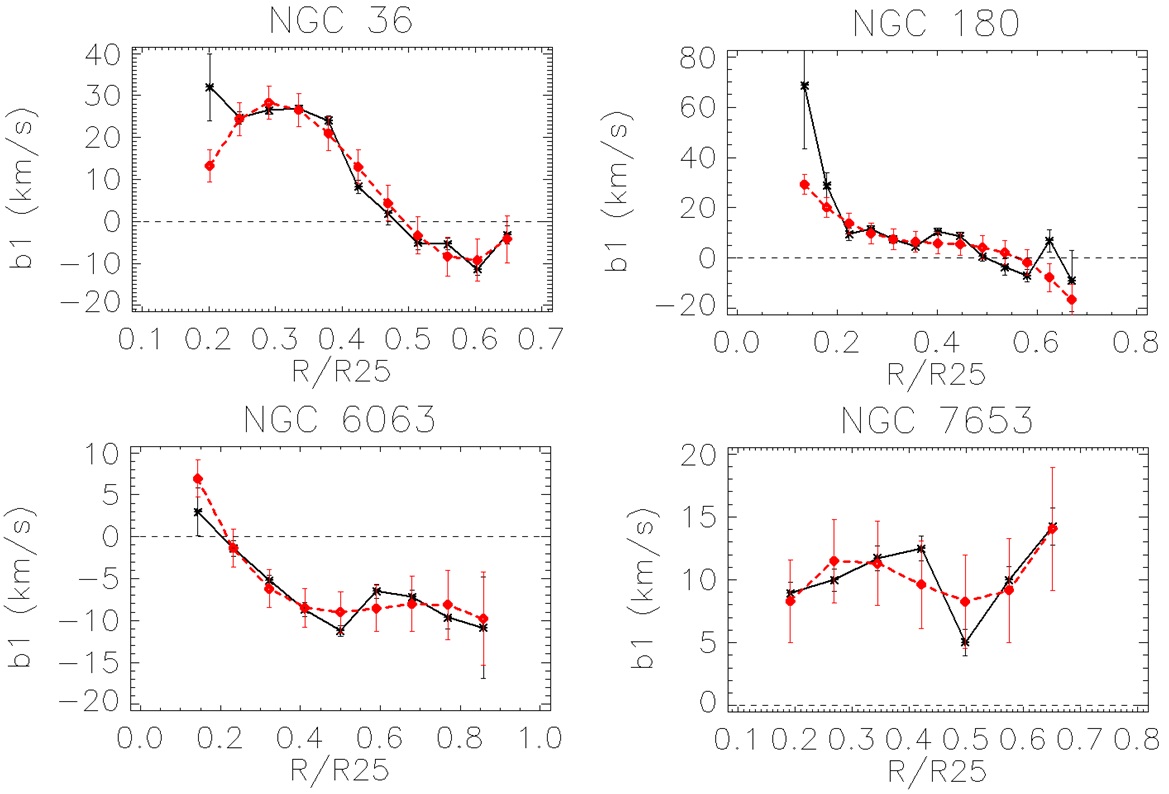}}
\caption{
 Fourier coefficient $b_1$ as a function of galactocentric distance
in our target galaxies (black curve).  The Fourier coefficient $b_1$ illustrates the radial change
of non-circular perturbations of the velocity caused by the radially symmetric motion $V_R$ and the second mode
($m=2$) of the spiral density wave (see Eq.~\ref{equation:Fourieir_Coef}).
The red curve is a model used to determine the radially symmetric motion $V_R$ and the amplitude of 
the velocity perturbation caused by the second mode of the spiral density wave.
}
\label{figure:b1}
\end{figure}

\subsection{Radially symmetric motions in the target galaxies obtained from velocity maps}
\label{sect:results_radial_motion}
%=============================================================================

The radially symmetric motion $V_R $ can be determined from the radial 
change of the first Fourier harmonics (black curves in Fig.~\ref{figure:b1}). 
Expression 3 of Eq.~\ref{equation:Fourieir_Coef} shows that the radial 
motion $V_R $ contributes to the Fourier coefficient ($b_1$) and mixes with 
the velocity perturbation from the second mode $m=2$ of the spiral density wave. 
First, we consider the barred galaxies NGC~36 and NGC~180. 
There are large radial velocities of $V_R > 30$ km~s$^{-1}$ (NGC~36)
and $V_R > 60$ km~s$^{-1}$ (NGC~160) 
in the inner parts of both barred galaxies. We assume that this is an
effect of the bar.  
The large error of coefficient $b_1$ in these points suggests that there is an explicit contribution of radial streaming 
velocities of the bar to the azimuthally averaged radial motion in NGC~36 and NGC~180.

Fig.~\ref{figure:b1} shows that there is a periodical component of the radial motion 
in the middle and outer parts of disc. 
The periodical component of the coefficient $b_1$ is shifted relative to 
the zero velocity line (dashed line in Fig.~\ref{figure:b1}). 
This may be an effect of the large-scale radial outflow/inflow 
motion outside of the bar region.
We approximated the radial change of the coefficient $b_1$ by a
polynomial function of the third degree (red curve).
As a sine or cosine shape of the variable term of the polynomial in barred galaxies 
corresponds rather to the spiral term than the bar term in Eq.~\ref{equation:Fourieir_Coef+Bar},
we made the simplest assumption that the variable term of the
polynomial describes 
the impact of the spirals. The constant term may be explained as an azimuthally 
averaged radial outflow/inflow motion, while the impact of the bar
term in the outer regions is negligible. 
This means that the coefficient $b_1(R)$ comprises generally the radial outflow/inflow 
motion $V_R$ and the radial component of the velocity perturbations from spirals. 
The relative small errors of the coefficient $b_1$ in the outer regions of 
the barred galaxies NGC~36 and NGC~180 may indicate that a contribution of 
radial streaming velocities in the bar to the azimuthally averaged radially 
symmetric motion in NGC~36 and NGC~180 is small outside the bar.
So we explain the radial change of the coefficient $b_1(R)$ with the help of 
the radial model with spiral perturbations (Eq.~\ref{equation:Fourieir_Coef}).
The curve can be interpreted as follows. The Fourier coefficient $b_1$ comprises 
a constant radially symmetric large-scale component $V_R$ and a periodical component of 
the non-circular motion caused by spiral perturbations outside the region with bar impact.
Table~\ref{table:result_V_R} lists the radial velocities in our target galaxies. 
Similar radial velocities were detected in NGC~2976 by \citet{spekkens2007}.

%++++++++++++++++++ Table1a   Results
\begin{table}
\caption[]{\label{table:result_V_R}
Radially symmetric motion in our target galaxies.
}
\begin{center}
\begin{tabular}{cccccccccccc} \hline \hline
Galaxy  & $V_R$ (km/s)$^a$ \\
\hline
NGC~36         & $9.1\pm 3.1$           \\
NGC~180      &  $7.0\pm 3.4$           \\
NGC~6063    &  $-4.6\pm 1.2$         \\
NGC~7653     &  $10.1\pm 1.4$         \\
\hline
\end{tabular}\\
\end{center}
\begin{flushleft}
$^a$ Since the side of the disk nearest to the observer disk along the minor 
axis is unknown, it is not possible to distinguish between inflow and outflow motion. 
\end{flushleft}
\end{table}

\subsection{Determination of the geometrical parameters of spiral arms 
from the velocity map}
\label{sect:spiral_arms_from_V_map}
%====================================================================

Despite the observational fact that spiral arms in a real galaxy never have a strict geometric form  
because other local structures may muddle their appearance, 
investigators always try to describe the large-scale spiral structure of galaxies 
using a model of regular geometric form. 
To minimize the influence of local structures, a large radial interval should be analyzed.
On the other hand, observations show that the pitch angle can vary along the radius 
(the change can exceed $20\%$ \citep{savchenko2013}). This suggests that a small radial
interval should be analysed.  
Therefore the number of annuli used to determine the pitch angle is smaller 
than the total number of annuli adopted when the Fourier coefficients were computed.

\begin{figure}
%\vspace{1.7mm}
%\resizebox{1.00\hsize}{!}{\includegraphics[angle=000]{Fig08.eps}}
\resizebox{1.00\hsize}{!}{\includegraphics[angle=000]{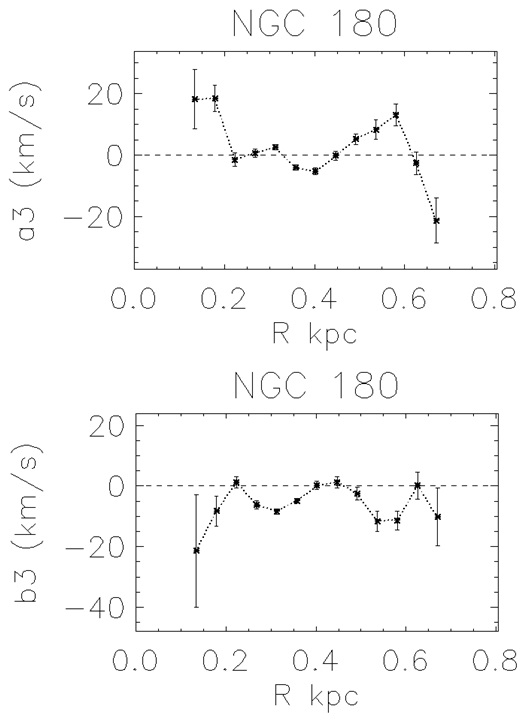}}
\caption{
The radial change of the Fourier coefficients $a_3$ and $b_3$ in NGC 180 shows 
that starting at about $0.2R_{25}$, the periodic component associated 
with the spiral velocity perturbation dominates the monotonically decreasing 
component of the bar term in the coefficients $a_3$ and $b_3$ from 
equation Eq.~\ref{equation:Fourieir_Coef+Bar}.
}
\label{figure:a3_b3_N180}
\end{figure}

In the current study, we adopt a logarithmic form of spiral arms
because it appears suitable and leads to a satisfactory fit.  We should
note that the procedure of the determination of the Fourier
coefficients does not assume that the spiral arms are logarithmic.
In earlier work,
spirals of different shapes were compared with observed spiral
structures \citep[][and references therein]{considere1982}.  It was
found \citep{considere1982} that the deviations of 
logarithmic and other shapes of spiral arms are smaller than the
dispersion in the positions of H\,{\sc ii} regions or other arm
tracers. 

 A logarithmic spiral is defined by the expression 
\begin{equation}
 R =R_0 e^{(\tan (\mu) \cdot \theta)}
\label{equation:spiral_pattern}
\end{equation}
where $R$ and $\theta$ are polar coordinates in the plane of the
galaxy, and $\mu$ is a pitch angle of the spiral arm. 

The change of  the coefficient $a_0$,  computed for the true $V_{sys}$,  with galactocentric
distance (see Fig.~\ref{figure:a0-Vsys} and Eq.~\ref{equation:first_mode}) provides a possibility to
 define graphically the pitch angle $\mu_1$ of the spiral arm.  The phase
difference between the maximum positive and the maximum negative
deviations from the (dashed) line of zero velocity is equal to $\pi$.  Fig.~\ref{figure:a0-Vsys}a
shows that the maximum positive deviation takes place at a
galactocentric distance $R_{max}/R_{25}=0.31$ and the maximum negative
deviation occurs at a galactocentric distance $R_{min}/R_{25}=0.58$ (marked with arrows)
for the galaxy NGC~36.  This results in a pitch angle of
$$\mu_1=\arctan\Bigl( \frac{1}{\pi 
}\ln\Bigl(\frac{R_{min}/R_{25}}{R_{max}/R_{25} }  \Bigr)\Bigr) 
\approx 11\degr$$ \\ 
where $R/R_{25}$ is a fractional radius (the galactocentric distance 
normalized to the disc's isophotal radius at 25 mag arcsec$^{-2}$).
 The graphical determination of the places with a given phase difference, 
e.g., $\pi$, for other galaxies is an ambiguous task. 
Assuming that the curve is a periodic function we determine the maximum 
positive and maximum negative deviations from the condition that the phase 
difference is equal to  $\pi$ (marked with arrows in Fig.~\ref{figure:a0-Vsys}). 
We find a pitch angle $\mu_1 \approx 17\degr$ for NGC~180, 
$\mu_1 \approx22\degr$ for NGC~6063, and $\mu_1 \approx 16\degr$ for NGC~7653.

The geometrical parameters of two-armed spirals $\mu_2$ and $R_{02}$
can be found from the radial change of the Fourier coefficients
$a_3(R)$ and $b_3(R)$ (see Eq.~\ref{equation:Fourieir_Coef}).  The
sixth and seventh expressions of Eq.~\ref{equation:Fourieir_Coef} give 
\begin{equation}
 \frac{a_3}{\sqrt{a_3^2+b_3^2} } = \sin \big(2\cot \mu_2 \ln{(R/R_{02})} -\mu_2 \big)
\label{equation:quotient_a0-K2}
\end{equation}

Eq.~(\ref{equation:quotient_a0-K2}) can be rewritten as a linear regression 
\begin{equation}
y = A + Bx   
\label{equation:lin_regression}
\end{equation}
with the notations 
\begin{eqnarray}
\label{equation:lin-regression_coeff}
x =\ln(R) \nonumber \\
y = \arcsin\Biggl(\frac{a_3}{\sqrt{a_3^2+b_3^2}}\Biggr)  \nonumber \\
A = \mu_2 -2\cot(\mu_2) \cdot \ln(R_{02}) \nonumber \\
B=2\cot(\mu_2) \nonumber
\end{eqnarray}
The estimation of the coefficients of the regression
(\ref{equation:lin_regression}) provides the geometrical parameters of
two-armed spirals, $\mu_2$ and $R_{02}$.
Table~\ref{table:result_kinematic} (Col.\ 6) shows estimations of the
pitch angle of $\mu_2$ of the two-armed spirals $\mu_2$ in our sample
using velocity field perturbations.  

We note the following:
NGC~36 has a relatively short bar, $R_{bar} \approx 0.14R_{25}$. 
We determine the pitch angle $\mu_2$ and characteristic radius $R_{02}$ from the coefficients $a_3$ and $b_3$ 
computed in the annuli outside the bar. So our radial model with spiral perturbations 
(Eq.~\ref{equation:v_light_sight} and Eq.~\ref{equation:Fourieir_Coef}) is applied. 

NGC~180 has a relatively long bar, $R_{bar} \approx 0.3R_{25}$.
The coefficients $a_3$ and $b_3$ used in the determination of the pitch angle $\mu_2$ and 
characteristic radius $R_{02}$ were computed for a number of annuli, 
some of which lie at the end of bar (from $0.23R_{25}$ to $0.3R_{25}$)
while the others 
are located outside the bar (from $0.3R_{25}$ to $0.4R_{25}$). 
Fig.~\ref{figure:a3_b3_N180} shows  that starting at about $0.2R_{25}$, the periodic term 
associated with the spiral velocity perturbation in the expressions (Eq.~\ref{equation:Fourieir_Coef+Bar}) 
for the coefficients $a_{3}$ and $b_{3}$ dominates over the monotonically decreasing term associated with the bar. 
 
The application of the radial model (Eq.~\ref{equation:v_light_sight} 
and Eq.~\ref{equation:Fourieir_Coef}) to the non-barred galaxies
NGC~6063 and NGC~7653 works well.

Since the pitch angle can vary along the radius (the change can exceed
20\% \citep{savchenko2013}), 
a limited interval of galactocentric distances $R$ was used in the determination of the pitch angle. 
This reduces the number of annuli, which lowers the accuracy of the pitch angle estimation.  
The intervals of galactocentric distances $R$ used for each tracer are given in Columns 7, 9, and 11 of the Table~\ref{table:result_kinematic}. 
These intervals do not cover the whole disc. 

\begin{figure}
%\vspace{1.7mm}
%\resizebox{1.00\hsize}{!}{\includegraphics[angle=000]{Fig09.eps}}
\resizebox{1.00\hsize}{!}{\includegraphics[angle=000]{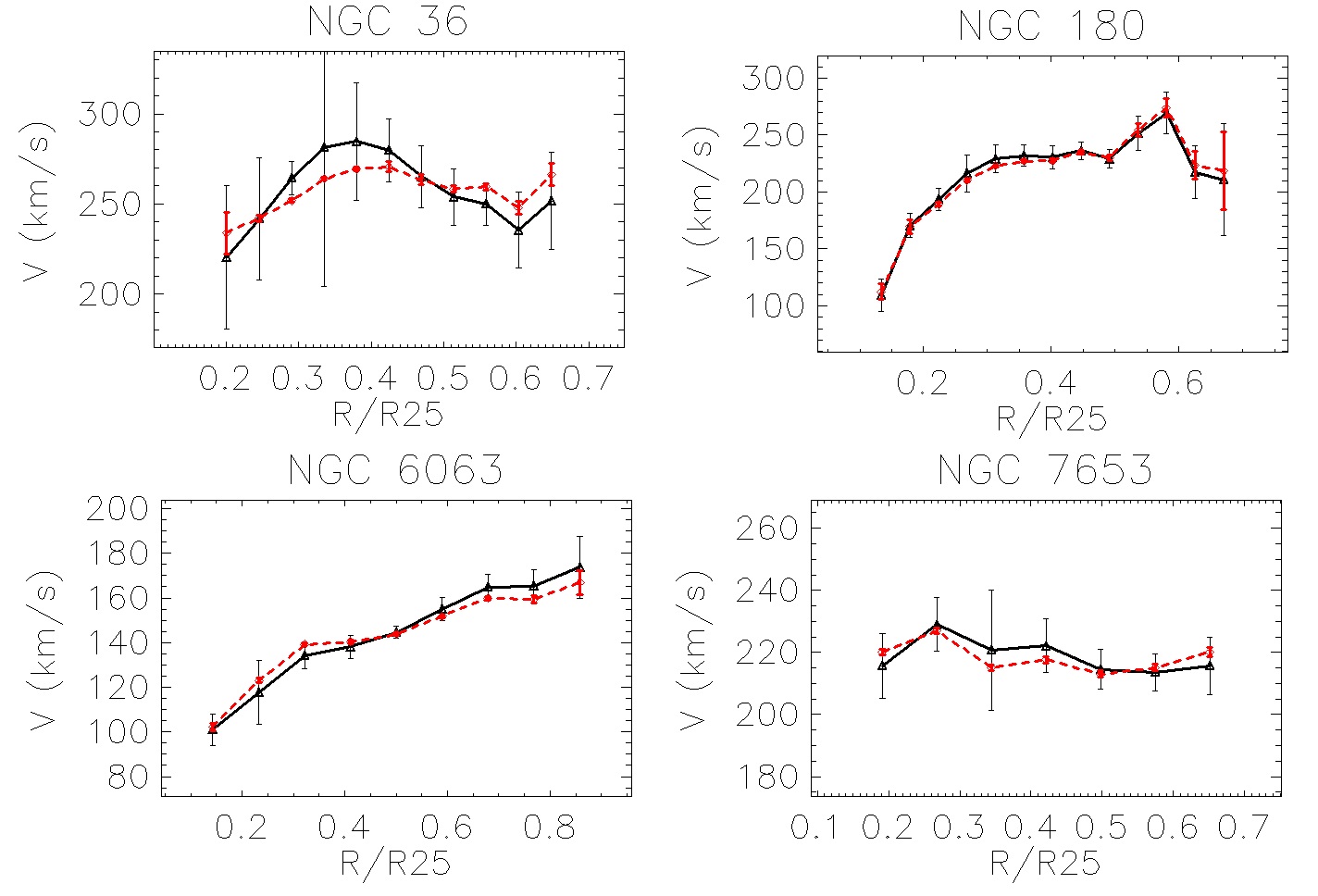}}
\caption{The rotation velocity $V_{\theta}(R)$ (solid black curve) and the  
value of the Fourier coefficient $a_1 (R)$ (dotted red curve) as a 
function of galactocentric distance for the target galaxies.
}
\label{figure:V_rot}
\end{figure}

%++++++++++++++++++ Table2   Results4
\begin{table*}
% \caption[]{\label{table:result_4}
\caption[]{\label{table:result_kinematic}
  The derived properties of our target galaxies (the position angle of 
  the major axis, $PA$, the galaxy inclination angle, $i$, the
  systemic velocity, $V_{gal}$, the maximum rotation velocity, $V_{max}$, 
  the pitch angle of the spiral arms, $\mu_2$, and the scaling factor 
  $R_{02}$).
}
\begin{center}
\begin{tabular}{cccccccccccc} \hline \hline
Galaxy & $PA^a$ &$i^a$ &   $V_{sys} $ & $V_{max} $    &$\mu_2^b$  &$\Delta R^b$ &$\mu_2^c$ & $\Delta R^c$ & $\mu_2^d$   & $\Delta R^d$  & $R_{02}$  \\
          &              &         &                      &                       &                &                     &              &                       &                 &                       &     \\
         &     deg    &  deg &        km/s       &km/s        &      deg     & $(R/R_{25})$&   deg     &  $(R/R_{25})$ &   deg   & $(R/R_{25})$ & $(R/R_{25})$ \\
   1     &     2       &   3    &    4    &          5            &       6       &         7          &      8       &          9           &       10      &         11          &  12  \\
\hline
NGC~36    & 14  &55 &6253$\pm$29  & 279   & 10  &  0.29 - 0.44  &  11 & 0.45 - 0.67 & 11 &  0.45 - 0.67 & $0.53\pm 0.09$   \\
NGC~180  &165 &46& 5405$\pm$14 & 277 & 18  &  0.23 - 0.37 &  17  & 0.24 - 0.38 & 18 & 0.23 - 0.45   & $0.29\pm 0.06$ \\
NGC~6063 &156&58 &2945$\pm$11 & 168 &  23  &  0.34 - 0.48 &  24  & 0.54 - 0.85 & 22 & 0.36 - 0.63  & $0.57\pm 0.15$  \\
NGC~7653 &165&31 &4356$\pm$21 & 225   & 14   &  0.52 - 0.65  &  15 & 0.31 - 0.47 & 12 & 0.31 - 0.44  & $0.43\pm 0.19$  \\
\hline
\end{tabular}\\
\end{center}
\begin{flushleft}
$(^a) $Uncertainties of the determination of $PA$ and $i$ are $\pm 2^o$.\\
$(^b)$Value estimated from the velocity field. \\ 
$(^c)$Value determined from the extinction map ratio $I_{H\alpha}/I_{H\beta}$. \\
$(^d)$Value is obtained from the oxygen abundance map. \\ 
The step size intervals of the galactocentric distances  $\Delta R$ used for the 
determination of the pitch angle of the spiral arms with different 
tracers are given in Columns 7, 9, and 11.
\end{flushleft}
\end{table*}

\subsection{Rotation curves of the target galaxies obtained from velocity maps}
\label{sect:results_kinematic}
%=============================================================================

The Fourier coefficients $a_1(R)$ and $b_1(R)$ of Eq.~(\ref{equation:Fourieir_Coef}) can be used to determine the
rotation curve of a galaxy. The second expression of Eq.~(\ref{equation:Fourieir_Coef}) can be rewritten as 
\begin{equation}
V_{\theta}(R) = a_1(R)-\frac{1}{2}(\hat u_2 - \hat v_2) \cdot\sin \big(2\cot \mu_2 \ln{(R/R_{02})} -\mu_2 \big)  .
\label{equation:rotation_curve}
\end{equation}
The spiral perturbations in Eq.~\ref{equation:rotation_curve} are
accounted for using the pitch 
angle $\mu_2$ obtained in Sec.~\ref{sect:spiral_arms_from_V_map} and the amplitude derived from  
the Fourier coefficient $b_1$. 
Neglecting the tangential bar term in the Eq.~\ref{equation:Fourieir_Coef} 
affects the estimation of the average rotation velocity and can result in  
a significant error of the rotation curve in the inner annuli for the
galaxy NGC~180 with its long bar. 
The rotation curve of the galaxy NGC~36 with a small bar is derived
outside the bar region.

Fig.~\ref{figure:V_rot} shows the rotation
velocity $V_{\theta}(R)$ (solid black curve) and the value of the Fourier
coefficient $a_1 (R)$ (dashed red curve) as a function of galactocentric
distance for the galaxies considered.  
We note that the error bar in the $a_1$ coefficient is substantially smaller than the error bar in $V_{\theta}$ 
except for a few cases, such as the last point of the rotation curve
in NGC~180, where the large error of $a_1$ corresponds 
to the large error of $V_{\theta}$.
Sources of large errors of rotation velocities $V_{\theta}(R)$ lie in
the uncertainties of the pitch angle $\mu_2$ caused by 
the complex structure of the spiral patterns and variations of the
pitch angle with galactocentric distance, as well as in uncertainties of the 
radial and tangential disturbances caused by the bar in barred galaxies.
A precise determination of the pure rotation curve remains a difficult task and depends on the applied method 
of the decomposition of the observed velocities.

\section{Spiral arms traced by abundance and extinction}
\label{sect:Z_Av_tracers}
%==============================================================

Here we examine the manifestation of spiral arms in the oxygen
abundance ($12+\log(O/H)$) and Balmer decrement
($I_{H\alpha}/I_{H\beta}$) distributions across the discs of our
galaxies.  Since the high-density (dust) clouds are concentrated along
the spiral arms one can expect that the extinction is higher in the
spiral arms than in the interarm regions. The star formation regions
are also concentrated in the spiral arms. As consequence, the Type II
supernova explosions responsible for the production and ejection of
the bulk of oxygen into the interstellar gas occur more often in
spiral arms than in the interarm regions.  If the rate of the
dissemination, dispersion, and mixing of the newly produced elements
in the gas is slow enough then one can expect that there is an oxygen
abundance enhancement in the spiral arms.

\subsection{Determination of spiral arms from the abundance (extinction) map}
%--------------------------------------------------------------

We assume that the oxygen abundance distribution in the interarm
regions of the disc, $I_{0}$, is axisymmetric and that the enhancement
of the oxygen abundance in the spiral arms, $I_{Sp}$, is small in
comparison to the oxygen abundance in the interarm region at a given
galactocentric distance.  Then the oxygen abundance distribution in
the disc, $I$, is given by the expression 
\begin{equation}
I(R,\theta) = I_0(R) + I_{Sp}(R,\theta) , \quad  \mid I_{Sp}\mid \ <<\ \mid I_0 \mid  .
\label{equation:sigma_global}
\end{equation}
The distribution of the enhancement of the oxygen abundance across the
spiral arms, $I_{Sp}$s, is assumed to be 
\begin{equation}
I_{Sp} = \hat I \cos \eta  .
\label{equation:sigma_spiral}
\end{equation}
According to Eq.~(\ref{equation:sigma_spiral}), the line of equal
spiral arm phase $\eta = const$ is a line of equal enhancement of the
oxygen abundance. It means that the centre of a spiral arm ($I_{Sp} =
\hat I$) is defined by the condition $\eta = 0$.  The values of $\eta
= \pm \pi / 2$ correspond to the outer and inner spiral arm borders
($I_{Sp}=0$).

Taking into account Eq.~(\ref{equation:eta_xi}) and
Eq.~(\ref{equation:sigma_spiral}) and grouping the sine and the cosine
terms of identical quantities of $\theta$, 
Eq.~(\ref{equation:sigma_global}) can be rewritten in the form
\begin{equation}
I(R,\theta) = a_0 +  a_1 \cos (\theta)+ b_1 \sin (\theta)+ a_2 \cos (2\theta)+ b_2 \sin (2\theta) 
\label{equation:sigma_spiral_2}
\end{equation}
where the coefficients $a_0(R), a_1(R), b_1(R), a_2(R)$, and 
$b_2(R)$ are 
\begin{eqnarray}
\label{equation:Fourieir_Coef_sigma}
a_0 = I_0(R) \nonumber \\
a_1=\hat I_1  \cos(\cot \mu_1 \ln{(R/R_{01})}   \nonumber\\
b_1=\hat I_1 \sin(\cot \mu_1 \ln{(R/R_{01})} \nonumber \\
a_2=\hat I_2  \cos(2\cot \mu_2 \ln{(R/R_{02})}  \\
b_2=\hat I_2  \sin(2\cot \mu_2 \ln{(R/R_{02})} \nonumber
\end{eqnarray}

The first mode $m = 1$  --  one-armed asymmetric structure --
contributes to the first Fourier harmonics of the azimuthal
distribution  of the  abundance at a given galactocentric distance $R$.
$ \hat I_1$ is the maximum enhancement of the oxygen abundance in the
centre of the one-armed spiral structure.  The second mode $m = 2$ --
a two-armed spiral -- contributes to the quantities multiplied by the
sines and cosines of the polar angle $2\theta$; i.e., to the second
Fourier harmonics.  $ \hat I_2$ is the maximum enhancement of the
oxygen abundance in the centre of the two-armed spiral structure.  The
coefficient $a_0(R) = I_0(R)$ is a function of the galactocentric
distance $R$ only and provides the radial abundance gradient in the
interarm regions (see Fig~\ref{figure:Gradient_OH}). 

The coefficients $a_0(R), a_1(R), b_1(R), a_2(R)$, and $b_2(R)$ were
obtained through the least-squares method using the abundance map
$I(R,\theta)$ for each target galaxy.  

Then we consider the radial change of the ratio 
\begin{equation}
 \frac{b_2}{\sqrt{a_2^2+b_2^2}} = \sin \bigl(2\cot \mu_2 \ln{(R/R_{02})} \bigr) .
\label{equation:quotient_b2-a2}
\end{equation}
With the definitions 
\begin{eqnarray}
\label{equation:lin-regression_coeff2}
x =\ln(R) \nonumber \\
y = \frac{1}{2}\arcsin \Biggl(\frac{b_2}{\sqrt{a_2^2+b_2^2} } \Biggr)  \nonumber \\
A =- \cot(\mu_2) \cdot \ln(R_{02}) \nonumber \\
B=\cot(\mu_2) \nonumber
\end{eqnarray}
Eq.~(\ref{equation:quotient_b2-a2}) can be rewritten as a linear
regression (Eq.~(\ref{equation:lin_regression})).  The coefficients
$A$ and $B$ of the linear regression are obtained through the 
least-squares method.  The pitch angle of a two-armed spiral, $\mu_2$, and
the constant $R_{02}$ are then defined by the obtained coefficients
$A$ and $B$.  Our results are presented in
Table~\ref{table:result_kinematic} (Column 10).

This method can be applied not only to the abundance maps, but to any
scalar field distributed across a galactic disc. In this study we also
apply it to the determination of spiral arms from the extinction maps
(Column 8 in Table~\ref{table:result_kinematic}).

\subsection{Metallicity gradients in discs and parameters of spiral arms  
derived from abundance maps}
\label{sect:Z_Av_results}
%---------------------------------------------------------------------------------

\begin{figure}
%\vspace{1.7mm}
%\resizebox{1.00\hsize}{!}{\includegraphics[angle=000]{Fig10.eps}}
\resizebox{1.00\hsize}{!}{\includegraphics[angle=000]{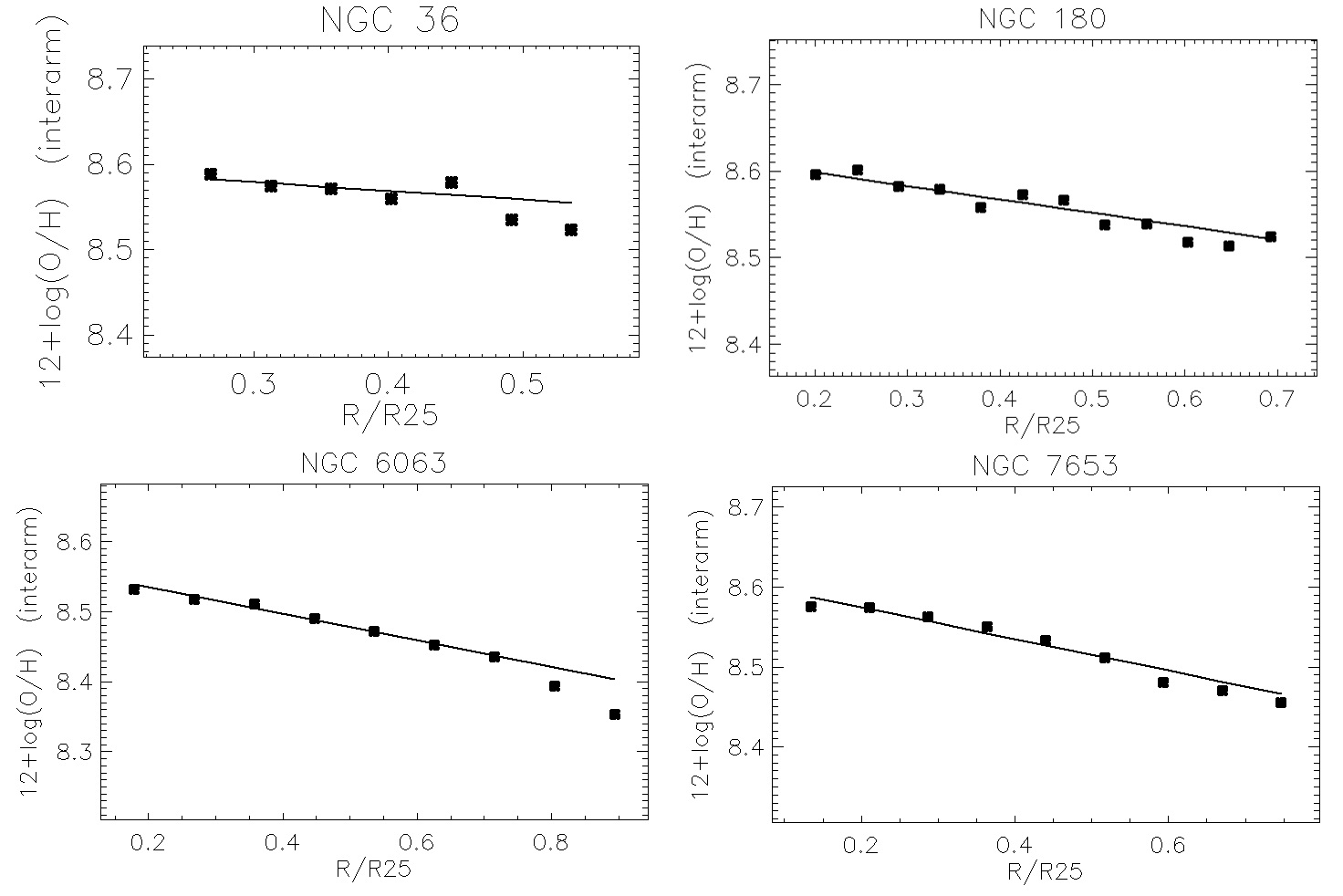}}
\caption{Radial abundance gradients in the interarm regions for our 
target galaxies.  The squares stand for the oxygen abundances at 
different radii. The line shows the best fit to those data.
}
\label{figure:Gradient_OH}
\end{figure}

We noted above that the coefficient $a_0(R) = I_0(R)$ gives the radial
abundance gradient in the interarm regions.
Fig~\ref{figure:Gradient_OH} shows the obtained radial oxygen
abundance gradients in the interarm regions in our four galaxies.  The
traditional form of the abundance gradient (linear regression) is
adopted: 
\begin{equation}
 12 + \log\big(O/H \big) = 12 + \log\big(O/H \big)_0 + grad \cdot \big(R/R_{25} \big) 
\label{equation:grad_OH}
\end{equation}
where $12+\log\big(O/H \big)_0$ is the extrapolated central oxygen
abundance and the $grad$ is the slope of the oxygen abundance
gradient expressed in terms of dex~$R_{25}^{-1}$. The $R/R_{25}$ is the
fractional radius (the galactocentric distance normalized to the disc
isophotal radius, $R_{25}$). 

The obtained central oxygen abundance and
the slope of the oxygen abundance gradient in the interarm regions of
our galaxies are listed in Table~\ref{table:result_OH} and shown in
Fig.~\ref{figure:Gradient_OH}.
  The slopes of the metallicity gradients expressed in terms of dex $R_{eff}^{-1}$ 
  are also obtained  using the values of the $R_{eff}$ from  \citet{Sanchez2014} 
  (Table~\ref{table:result_OH}).

\begin{figure}
%\vspace{1.7mm}
%\resizebox{1.00\hsize}{!}{\includegraphics[angle=000]{Fig11.eps}}
\resizebox{1.00\hsize}{!}{\includegraphics[angle=000]{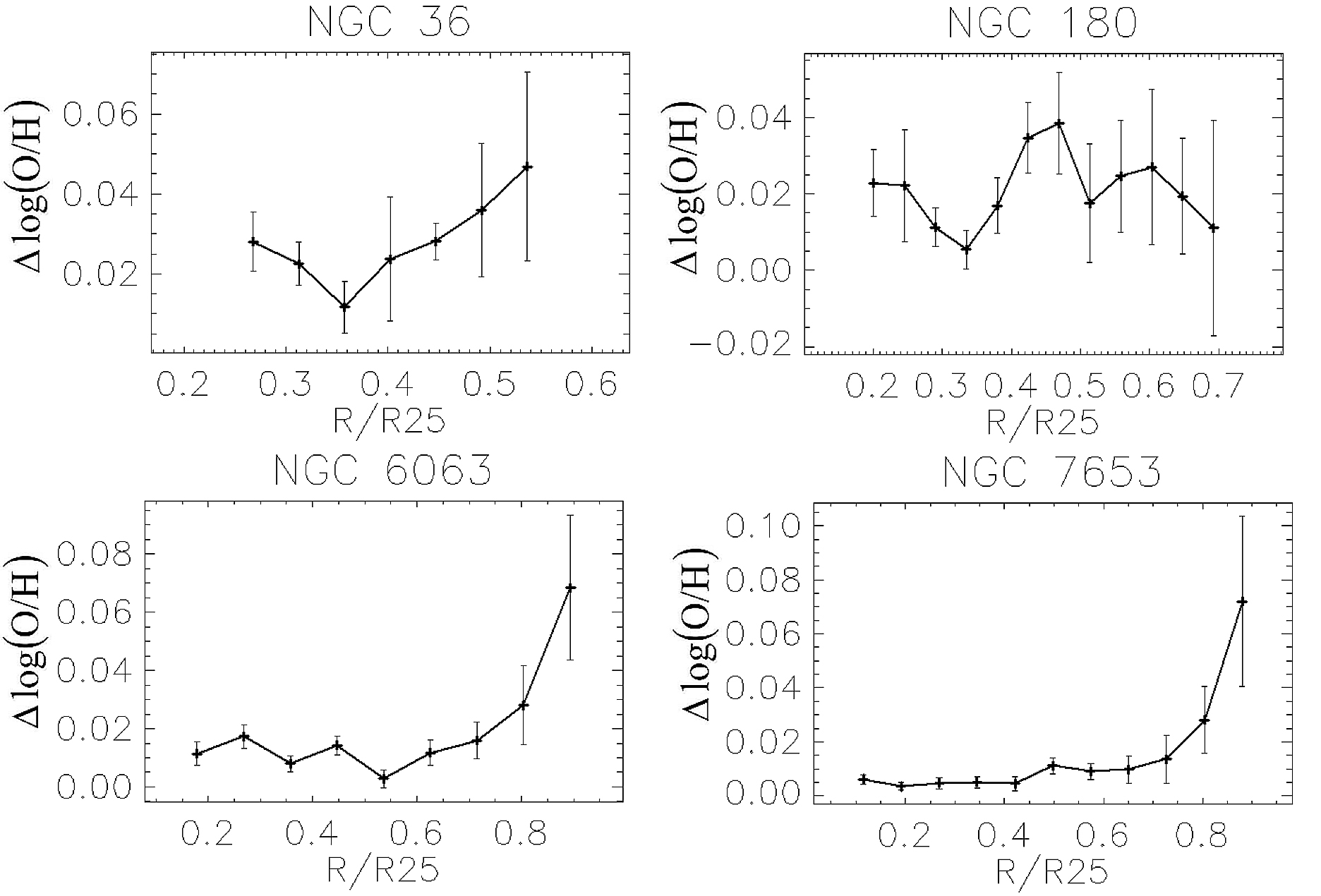}}
\caption{The enhancement of the oxygen abundance $\Delta \log$(O/H) 
in the spiral arms as compared to the interarm regions as a function 
of galactocentric distance for the target galaxies.
%$\log\frac{(O/H)_{Spiral}}{(O/H)_{Interarm}}$
}
\label{figure:relative_OH}
\end{figure}

Fig~\ref{figure:relative_OH} shows the enhancement of the oxygen
abundance in the spiral arms as compared to the mean (azimuthally
averaged) abundance in the interarm regions at a given galactocentric
distance as a function of galactocentric distance for the galaxies
considered.  The enhancement caused by the second mode ($m=2$) of the
spiral density waves is considered.  Inspection of
Fig~\ref{figure:relative_OH} shows that the enhancement of the oxygen
abundance in the spiral arms is small and increases outwards from
$\sim 2$ to $\sim 8$ per cent.

The comparison between the parameters of the radial distributions of
the oxygen abundance in the interarm regions (the extrapolated central
oxygen abundance and the slope of the oxygen abundance gradient)
listed in Table~\ref{table:result_OH} and the parameters of the radial
abundance distribution for the whole discs (spiral arms and interarm
regions) reported in \citet{Zinchenko2016} shows that these two sets
of values are close to each other. This suggests that the influence of
the spiral arms on the radial abundance gradient in galactic discs is
small. 

It should be noted that the location of the spiral arms can be
established from the abundance map not for the whole disc but only for
some range of  galactocentric distances. These ranges are given in
Table~\ref{table:result_kinematic} (Column 11). The spiral arms start
at some dstance from the center and therefore the innermost part of
the galaxies is excluded from our considerations. The spiral arms at
the peripheral outer regions of our galaxies cannot be established
because of the poor statistics of the observed points.

%++++++++++++++++++ Table3   Results
\begin{table}
\caption[]{\label{table:result_OH}
The extrapolated central oxygen abundance and the slope of the 
oxygen abundance gradient
}
\begin{center}
\begin{tabular}{cccccccccccc} \hline \hline
Galaxy  & $12+\log\big(O/H \big)_0$ &$grad $  & $grad $ & $R_{25}/R_{eff}$ \\
         &                                           & $  dex/R_{25} $  & $  dex/R_{eff}  $  &      \\
      1 &     2                                    &      3       & 4   & 5   \\
\hline
NGC~36         & $8.61\pm 0.01$       & $ -0.10 \pm 0.02$   & $ -0.07$     & 1.53   \\
NGC~180      &  $8.63\pm 0.01$      &$ -0.16 \pm 0.02 $    & $-0.09$      & 1.77  \\
NGC~6063    &  $8.573\pm 0.002$  &$ -0.189\pm 0.005 $  & $-0.09$  & 2.02    \\
NGC~7653     &  $8.614\pm 0.001$  &$ -0.196\pm 0.003 $ & $-0.09$  & 2.15     \\
\hline
\end{tabular}\\
\end{center}
\begin{flushleft}

\end{flushleft}
\end{table}

\begin{figure}
%\vspace{1.7mm}
%\resizebox{1.00\hsize}{!}{\includegraphics[angle=000]{Fig12.eps}}
\resizebox{1.00\hsize}{!}{\includegraphics[angle=000]{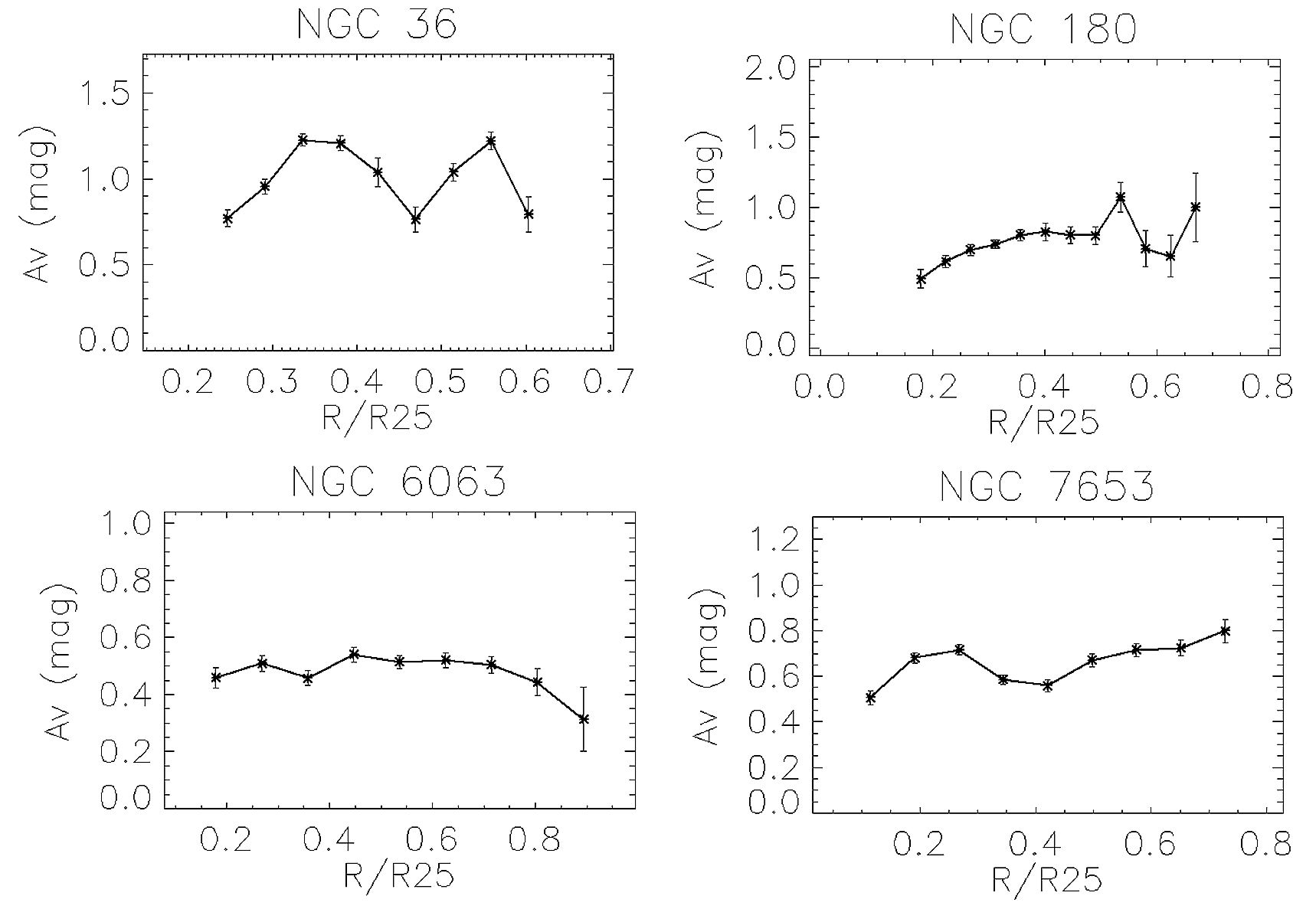}}
\caption{The visual extinction $A_{V}$ as a function of galactocentric
distance in our target galaxies.  The points mark the azimuthally
averaged values of extinction. The error bars show the scatter.
}
\label{figure:Av}
\end{figure}

\subsection{Spiral arms in our galaxies derived from extinction maps}
\label{sect:Radial_Av}
%--------------------------------------------------------------------

The method described above and used for the analysis of the abundance
map to trace the spiral arms can be also applied to the extinction
map.  Here we search for spiral arms in our target galaxies using the
Balmer decrement and succeeded in uncovering the spiral arms in each
target galaxy. 

Fig~\ref{figure:Av} shows the azimuthally averaged value of the visual
extinction $A_V$ in the interarm regions as a function of
galactocentric distance in our target galaxies. Inspection of
Fig~\ref{figure:Av} shows that there is no common systematic change of the
extinction along the radial extent of the disc; instead the extinction
is constant within the uncertainties.  

\begin{figure}
%\vspace{1.7mm}
%\resizebox{1.00\hsize}{!}{\includegraphics[angle=000]{Fig13.eps}}
\resizebox{1.00\hsize}{!}{\includegraphics[angle=000]{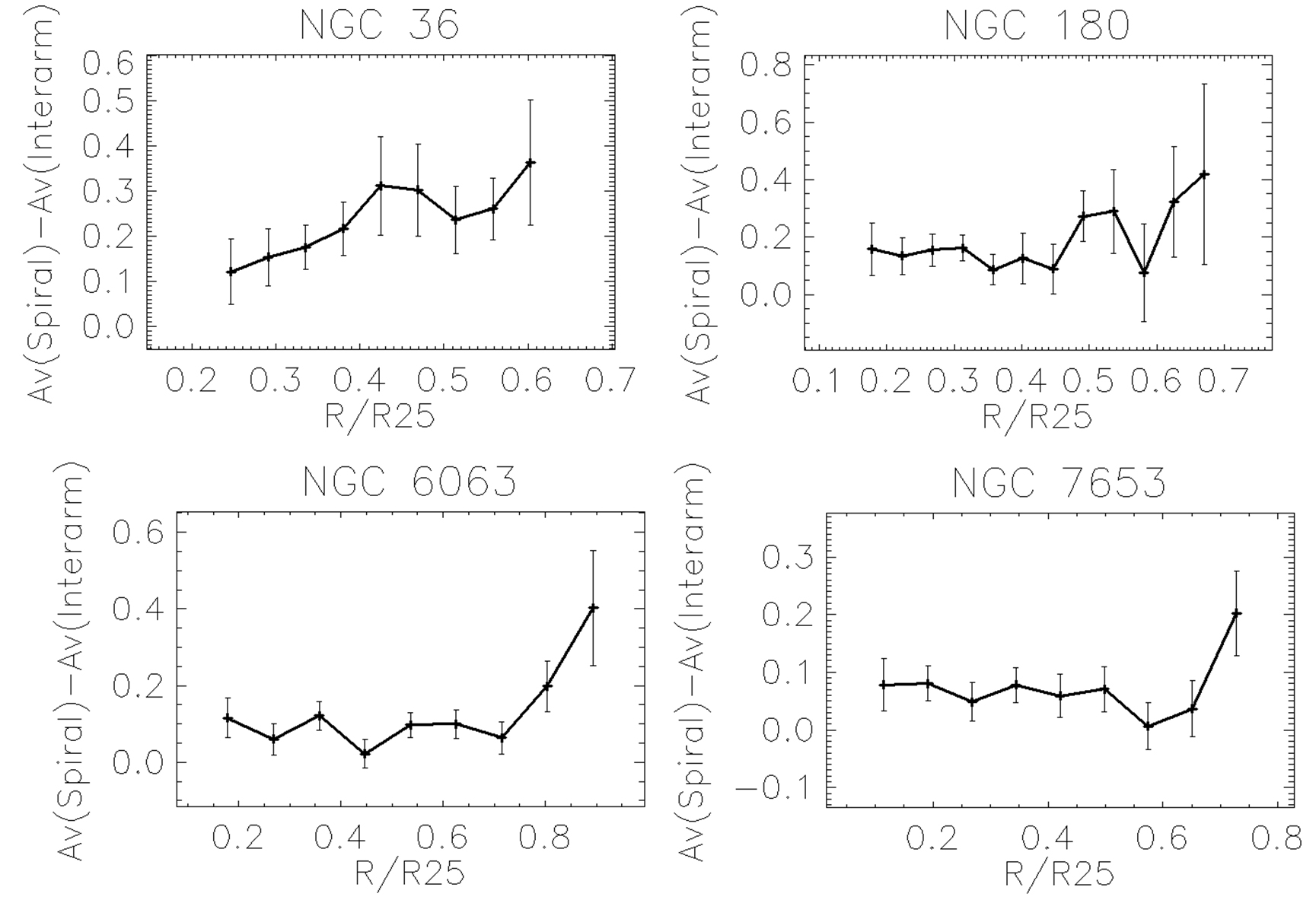}}
\caption{The difference between the total visual extinction in the 
spiral arms, $A_V(Spiral)$, and the mean extinction in the interarm 
regions, $A_V(Interarm)$, as a function of galactocentric distance.
}
\label{figure:relative_Av}
\end{figure}

Fig~\ref{figure:relative_Av} shows the difference between the
extinction in the spiral arms, $A_V(Spiral)$, and the mean extinction
in the interarm regions, $A_V(Interarm)$, as a function of
galactocentric distance.  The difference  $A_V(Spiral) -
A_V(Interarm)$ increases monotonically outwards.

\begin{figure*}
%\resizebox{1.00\hsize}{!}{\includegraphics[angle=000]{Fig14.eps}}
\resizebox{0.90\hsize}{!}{\includegraphics[angle=000]{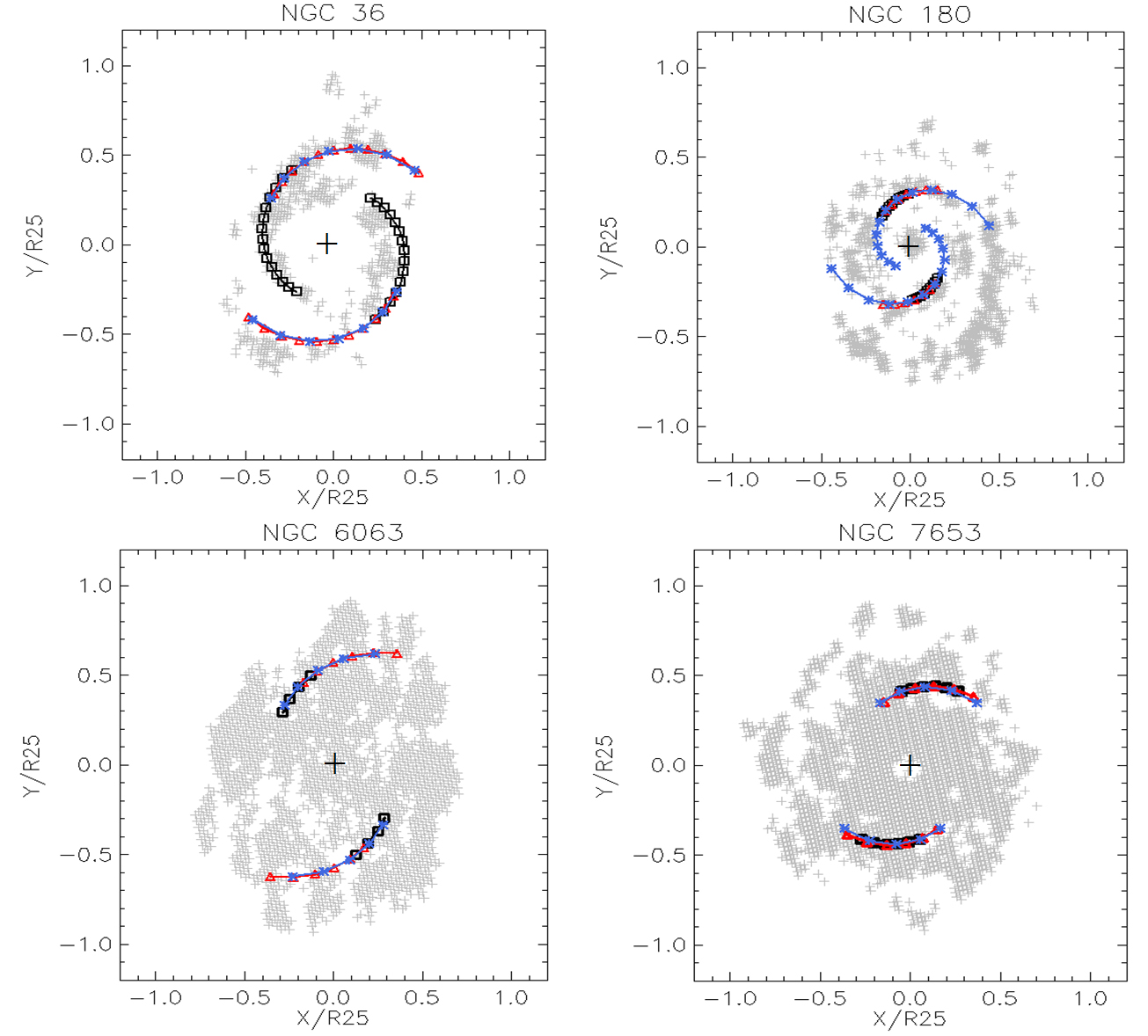}}
\caption{Comparison between spiral arms derived from the analysis of 
the velocity map (black squares), the extinction map (red triangles), 
and the oxygen abundance map (blue stars).  The small grey plus 
signs mark the areas (pixels) with measured oxygen abundances. The 
large dark plus sign marks the position of the centre of the galaxy.
Red rectangle shows the orientation and size of the bar in NGC~36 and NGC~180 (upper panels) 
}
\label{figure:maps}
\end{figure*}

\section{Discussion}
\label{sect:discussion}
%========================

Fig.~\ref{figure:maps} shows the spiral arms obtained from the maps of
different tracers in the target galaxies.  The spiral arms in
Fig.~\ref{figure:maps} are shown within the range of galactocentric
distances where they were detected (see
Table~\ref{table:result_kinematic}). 

Inspection of Fig.~\ref{figure:maps} (compare also the values of the
pitch angle listed in Columns 6, 8, and 10 of
Table~\ref{table:result_kinematic}) shows that the parameters of the
spiral arms derived from maps of different tracers are in good
agreement with each other.  This is not surprising because of all
three tracers are associated with ionized gas clouds.  Since the
high-density (dust) clouds are concentrated in the spiral arms one may
expect a higher extinction in the spiral arms than in the interarm
regions.  The type II supernovae responsible for the oxygen enrichment
of the interstellar gas are more often located in the spiral arms than
in the interarm regions.  Thus the spiral arms derived from the maps
of different tracers are in line with the gravitational density wave
nature of the spiral arms and the theory of the star formation.
  A similar conclusion has been reached by \citet{SanchezMenguiano2016b} on the
  base of the analysis of the VLT/MUSE data of the galaxy NGC~6754.

The reliability and accuracy of the parameters of the spiral arms
obtained for each tracer is rather low because of the small amplitude
of the disturbance of the tracer caused by the spiral arms and the
relatively low spatial resolution of the observations.  Moreover, a
detailed photometric study and measurements of spiral arm pitch angles
in 50 grand-design spiral galaxies \citep{savchenko2013} shows that
about 2/3 of these galaxies demonstrate pitch angle variations exceeding 20
per cent.  This also influences the accuracy of the definition of
pitch angles.  The fact that the parameters of the spiral arms derived
from the maps of our different tracers are close to each other
supports the reliability of the obtained geometrical parameters of the
spiral patterns.

\citet{SanchezBlazquez2014} noted that the non-linear coupling between 
the bar and the spiral arms is an efficient mechanism for producing radial 
migrations across significant distances within discs.
The process of radial migration should flatten the stellar metallicity gradient 
with time and, therefore, one can expect flatter stellar metallicity gradients in barred galaxies.
  It is interesting to note that the metallicity gradients expressed in terms
  of dex $R_{25}^{-1}$ in the barred galaxies NGC~36 and NGC~180 are flatter than
  in the unbarred galaxies NGC~6063 and NGC~7653. 
  However, if the gradient is expressed in terms of dex $R_{eff}^{-1}$  then
  there is no difference between the gradients in barred and unbarred galaxies
  (Table~\ref{table:result_OH}).
  This agrees with the conclusion of \citet{Sanchez2014} and \citet{SanchezBlazquez2014} that
  there is no difference in the slope of the metallicity gradients expressed in terms
  of dex $R_{eff}^{-1}$ between barred and unbarred galaxies. 

We found that the early-type spiral galaxies (NGC~36 and NGC~7653)
demonstrate more tightly wound arms in comparison with the late-type
spiral galaxies (NGC~180 and NGC~6063) as one would expect based on
their morphological types. This is in agreement with previous results
\citep[e.g.,][]{Kennicutt1981,savchenko2013}. 

It should be noted that \citet{vogt2017} detected the presence of
rapid abundance variations in the spiral galaxy HGC 91c based on the
Multi Unit Spectroscopic Explorer (MUSE) observations. These
variations can be separated in two distinct types, namely
sub-kpc-scale variations associated with individual star-forming
regions and kpc-scale variations correlated with the spiral structure
of the galaxy. The authors conclude that the kpc-scale variations thus
provide observational evidence that ISM enrichment is preferentially
occurring along the spiral structure of HCG 91c, and less easily
across the spiral arms.

It is known that the maximum rotation velocity, $V_{max}$, of disc
galaxies correlates with the absolute blue magnitude of the galaxy
$M_B$ and its morphological type \citep*{Rubin1980, Rubin1982,
Rubin1983, burstein1985}. 
The values of the maximum rotation velocity $V_{max}$ derived in our
current study (Column 5 in Table~\ref{table:result_kinematic}) are in
agreement with the characteristic maximum rotation velocities for the
galaxies of similar luminosity and morphological type.

We discussed in Sec.~\ref{sect:results_kinematic} a possible impact of the bar 
on the determination of the rotation curve $V_{\theta}(R)$. 
Through the analysis of the Fourier coefficient $b_1$ we found large 
radial velocities $V_R \approx (30-60)$ km~s$^{-1}$ in the inner parts of 
both barred galaxies, while in outer part of the galaxies the radially 
symmetric large-scale motion is lower; $V_R \approx 10$ km~s$^{-1}$. 

  The position angles of the major axis of a galaxy obtained here from
  the velocity field agree  with the position angles determined
  from the analysis of the surface brightness distributions in \citet{Zinchenko2016}.
  That is in line with the result of \citet{BarreraBallesteros2014},  
  who did not find statistically significant differences between the kinematic $PA$ and
  the global photometric orientation of the galaxy, including barred and unbarred objects.

Fig.~\ref{figure:a0-Vsys} shows a variation of the zeroth coefficient
$a_0$  with increasing galactocentric distance in our target galaxies
(the true systemic velocity is subtracted from the measured velocity of each spaxel). 
Such a dependence of the zeroth coefficient on the galactocentric
distance was also detected by \cite{carignan1990} in the spiral galaxy
NGC~6946 and by \cite{ sakhibov2004} in NGC~628.  \cite{carignan1990}
assumed that the coefficient  $a_0$ comprises just the systemic
velocity of the galaxy as awhole, $V_{gal}$, and explained the radial dependence
of the zeroth coefficient  by the periodic variation of the
inclination of the galaxy disc from $i =33^{\circ}$ up to $i
=43^{\circ}$.  \cite{sakhibov2004} suggested that the radial
dependence of the zeroth coefficient $a_0$ in NGC 628 can be explained
with the first mode of a density wave, which makes a contribution in
zeroth harmonics.
The superposition of the first and second modes can result in asymmetric 
spiral structure, which is apparent in the maps of the metallicity distribution 
of NGC~180 and NGC~7653  (Fig.~\ref{figure:maps}).

  \citet{Sanchez2015} and \citet{SanchezMenguiano2016b}
  considered the azimuthal distribution of the residuals of
  the oxygen abundance for the individual H\,{\sc ii} regions in the
  galaxy NGC~6754 after subtracting the average radial gradient.
  They noted that the strongest residuals correspond to the spiral arms.   
  The mean absolute value of the strongest residuals is below $\sim$0.05 dex
  and changes with galactocentric distance.
  We found a similar picture for the enhancements of the oxygen abundance in the
  spiral arms as shown in Fig.~\ref{figure:relative_OH}.

  \citet{SanchezMenguiano2017} carried out a study of the arm
  and interarm abundance distributions in a sample of 63 CALIFA spiral galaxies. 
  They found that the differences between the gas abundances of spiral
  arms and interarm regions are small and statistically significant only for
  flocculent and barred galaxies. The small values of the enhancement of the oxygen
  abundance in the spiral arms obtained here for four galaxies is in
  agreement with their results.

\section{Conclusions}
\label{sect:conclusions}

We constructed maps of the observed velocity of the ionized gas, the
oxygen abundance, and the extinction (Balmer decrement) in the discs
of the four spiral galaxies NGC~36, NGC~180, NGC~6063, and NGC~7653
based on integral field  spectroscopy obtained by the Calar Alto
Legacy Integral Field Area (CALIFA) survey. The maps were used to
search for spiral arms in the discs of these galaxies via Fourier
analysis.  We considered non-circular motions, the enhancement of the
oxygen abundance, and increased extinction as tracers of spiral arms.

The parameters of the spiral structure were obtained in all four
galaxies considered and for each tracer.  The pitch angles of the
spiral arms obtained from the different tracers are close to each
other.  This strengthens the reliability of the obtained geometrical
parameters of the spiral patterns although the accuracy of
determination of these parameters for each tracer separately is low.
The systemic velocity, inclination, position angle of the major axis,
and rotation curve were also obtained for each target galaxy using the
Fourier analysis of the two-dimensional velocity map.

The enhancement of the oxygen abundance in the spiral arms as compared
to the abundance in the interarm regions at a given galactocentric
distance is small, within a few per cent.  This evidences that the
influence of the spiral arms on the present-day radial abundance
gradient in the discs of spiral galaxies is small and can be neglected
in many tasks.  There is no regular change of the extinction with
galactocentric distance in the studied galaxies.

\section*{Acknowledgments}

We are very grateful to the referee for  his/her constructive comments and recommendations, which significantly improved the article.  \\
I.~A.~Z., L.~S.~P., E.~K.~G., and A.~J.\ acknowledge support within 
the framework of Sonderforschungsbereich SFB 881 on ``The Milky Way 
System'' (especially via subprojects A5 and A6). SFB 881 is funded by 
the German Research Foundation (DFG). \\ 
I.~A.~Z.\ and L.~S.~P.\ and thank for the hospitality of the
Astronomisches Rechen-Institut at Heidelberg University, where part of
this investigation was carried out. \\
I.~A.~Z.\ acknowledges the support of the Volkswagen Foundation 
under the Trilateral Partnerships grant No.\ 90411. \\
This study uses data provided by the Calar Alto Legacy Integral Field Area 
(CALIFA) survey (http://califa.caha.es/). 
Based on observations collected at the Centro Astron\'{o}mico Hispano 
Alem\'{a}n (CAHA) at Calar Alto, operated jointly by the 
Max-Planck-Institut f\"{u}r Astronomie and the Instituto de 
Astrof\'{i}sica de Andaluc\'{i}a (CSIC). \\ 
The authors acknowledge the usage of the HyperLeda data base 
(http://leda.univ-lyon1.fr), the NASA/IPAC Extragalactic Database 
(http://ned.ipac.caltech.edu).

\end{document}